\documentclass[twocolumn]{aastex631}

\usepackage{tablefootnote}

\shortauthors{Panzera et al.}

\begin{document}

\title{Accretion Properties of the Young Brown Dwarf 2MASS J08440915-7833457}

\correspondingauthor{Toni V. Panzera}
\email{tvp3@rice.edu}

\author{Toni V. Panzera}
\affiliation{Department of Physics and Astronomy, Rice University, 6100 Main Street, Houston, TX 77005, USA}






\author{Laura S. Flagg}
\affiliation{Department of Physics and Astronomy, Johns-Hopkins University, 3400 N. Charles Street, Baltimore, MD 21218, USA}

\author{Margaret A. Mueller}
\affiliation{Department of Physics and Astronomy, Rice University, 6100 Main Street, Houston, TX 77005, USA}

\author{Christopher M. Johns-Krull}
\affiliation{Department of Physics and Astronomy, Rice University, 6100 Main Street, Houston, TX 77005, USA}

\author{Gregory J. Herczeg}
\affiliation{Department of Astronomy, Peking University, Yiheyuan Lu 5, Haidian Qu, 100871 Beijing, People’s Republic of China}
\affiliation{Kavli Institute for Astronomy and Astrophysics, Peking University, Yiheyuan Lu 5, Haidian Qu, 100871 Beijing, People’s Republic of China}





\begin{abstract}

We present \textit{HST}-COS FUV and -STIS optical observations towards the young accreting brown dwarf 2MASS-J08440915-7833457 (J0844) from the ULLYSES DDT Program. We analyse hot FUV lines such C \textsc{\footnotesize{IV}}, Si \textsc{\footnotesize{IV}}, and N \textsc{\footnotesize{V}}, as well as fluorescent emission from H$_2$. Despite evidence for accretion, the C \textsc{\footnotesize{IV}} line profiles are narrower than in typical classical T Tauri stars (CTTSs), resembling weak-lined T Tauri stars more closely. Additionally, the C \textsc{\footnotesize{IV}} integrated line flux does not follow the level expected of an accreting object in the magnetically saturated regime. However, comparing J0844 to appropriate low mass analogs, J0844 does show excess C \textsc{\footnotesize{IV}} emission characteristic of accretion, suggesting the magnetic saturation level may need to be redefined for the lowest mass objects. The C \textsc{\footnotesize{IV}}/Si \textsc{\footnotesize{IV}} emission line ratio is found to be 20, which is higher than most CTTSs, with a few exceptions (e.g., TW Hya). We fit the STIS optical spectrum to calculate an accretion rate, which we find to be $4.2 \times 10^{-11}$$M_\odot$ yr$^{-1}$. The accretion rate found based on the empirical $L_\mathrm{CIV}$--$\dot{M}_\mathrm{acc}$ relationship is two orders of magnitude higher, suggesting this relationship may not hold at the lowest masses. We find the H$_2$ emission appears to originate within the co-rotation radius, pointing to either disc truncation well inside the co-rotation radius or additional sources of H$_2$ emission that we do not consider (e.g., from the accretion flow itself). These data provide an extension of our current understanding of accretion and inner disc conditions to the relatively unexplored lowest mass regime.

\end{abstract}



\section{Introduction} \label{sec:intro}

Young stars emerge from their natal nebulae and become optically visible when they enter the very active T Tauri phase of their evolution, after about one million years. T Tauri stars (TTSs) are magnetically active (\citealt{jk2007, hartmann_accretion_2016}) and are often surrounded by accretion discs (\citealt{hartmann_optical_1990}) which are influenced by both inflows and outflows. As such, accretion physics plays an integral role in the evolution of these pre-main sequence (PMS) stars (\citealt{mercer-smith_formation_1984, hartmann_optical_1990}). 

The properties of accreting TTSs (generally referred to as classical TTSs; CTTSs) have been relatively well-studied (\citealt{hartmann_why_2006, herczeg_twenty-five_2023, donati_classical_2024}), and it is well-established that accreting material flows from the inner truncation boundary of the disc along stellar magnetic field lines, reaching near free-fall velocity close to the stellar surface where an accretion shock forms (see review by \citealt{hartmann_accretion_2016} and references therein). The shocked gas heats up to about a million degrees, and subsequently cools and becomes more dense as it settles onto the surface of the star (\citealt{calvetgullbring1998}). The presence of this hot accreting material produces several broad emission lines in the far-ultraviolet (FUV), such as C \textsc{\footnotesize{IV}}, N \textsc{\footnotesize{V}} and Si \textsc{\footnotesize{IV}} which probe gas at approximately $10^5$ K (\citealt{ardila_hot_2013, skinner_hst_2022, than2024}).  As a result, the strength and shape of these lines is important in our understanding of the accretion process. 

However, the production of hot FUV lines such as C \textsc{\footnotesize{IV}} cannot always be attributed entirely to accretion. It is well-known that magnetically active stars including our Sun also produce C \textsc{\footnotesize{IV}} emission through chromospheric and transition region activity (e.g., \citealt{gomez_de_castro_extended_2012}), and the magnetic fields of TTSs can be on the order of a few kG (\citealt{guenther_measurements_nodate, jk1999, jk2007}), suggesting the potential for significant magnetic activity related emission in these lines. Therefore, it is important to distinguish the level of emission from accretion and magnetic activity. To this end, \cite{jk_2000} showed that non-accreting TTSs (weak-line TTSs; WTTSs) show levels of C \textsc{\footnotesize{IV}} emission at or below the expected magnetic activity saturation level, while CTTSs are at this level or above (see also \citealt{yang_far-ultraviolet_2012}). Additionally, this excess emission in CTTSs is well-correlated with the accretion rate (\citealt{jk_2000}).

More recently, accretion onto forming giant planets has been actively explored (e.g., PDS 70b; \citealt{hashimoto_accretion_2020}), and it has been suggested that this accretion process occurs in a similar way as it does in stars. Aside from C \textsc{\footnotesize{IV}}, another accretion signature is strong H$\alpha$ emission which probes gas at about $10^4$ K (\citealt{calvet_balmer_1992, muzerolle_accretion_2003, tsilia_photometric_2023, lin_mass_2023}), which has been observed in planetary mass objects (\citealt{eriksson_strong_2020,haffert_two_2019,zhou_accretion_2014,zhou_hstwfc3_2022}). However, the relationship between H$\alpha$ luminosity and accretion rate is more uncertain at these low masses. For example, \cite{huelamo_searching_2022} and \cite{than2024} showed that accretion estimates based on the $L_{\mathrm{H}\alpha}$--$\dot{M}_\mathrm{acc}$ relationship yield mass accretion rates one to two orders of magnitude lower than what is predicted by planetary models (e.g., \citealt{aoyama_comparison_2021}). In a similar vein, \cite{zhou_accretion_2014, zhou2021} found that the ratio of hydrogen line luminosity to accretion luminosity was higher for planetary-mass objects compared to CTTSs, indicating increased efficiency in hydrogen line production in very low mass objects. These differences may point towards a difference in the accretion flow structure itself as the mass of the central object decreases, indicating that existing stellar accretion relationships cannot simply be extrapolated down to the lowest mass objects.

In addition to analysing the accretion flows, UV spectroscopy can probe the structure of the inner disc. Usually, emission radii of molecules such as CO are determined based on infrared (IR) observations (\citealt{salyk_co_2011}), since the composition of the disc is mostly molecular at these temperatures. There is also the opportunity to study molecular disc emission in the UV from the electronic transitions of H$_2$: H$_2$ UV emission lines are commonly observed in CTTS systems, caused primarily by Ly$\alpha$ pumping (\citealt{herczeg_farultraviolet_2002, france_radial_2023}) and have similarly been used to calculate H$_2$ emission radii in discs (\citealt{france_hubble_2012, france_radial_2023, gangi_penellope_2023}).

The brown dwarf regime offers an opportunity to bridge the gap between our understanding of stellar and planetary accretion flows and inner disc properties. By studying these low-mass objects, we can test whether our current understanding of the accretion process in stars can simply be extended towards the planetary regime, as well as map out if, and how, accretion changes as the mass of the accreting object decreases. 

It is well-known that many brown dwarfs go through a CTTS phase where they accrete material from a circumstellar disc (e.g., \citealt{muzerolle_detection_2000, megeath_spitzerirac_nodate}). Until now, these accretion estimates have generally been based on optical data (usually H$\alpha$ luminosity; \citealt{muzerolle_accretion_2003, wu_monitoring_2023}), though a few have been explored in terms of their FUV emission (e.g., \citealt{france_metal_2010, yang_far-ultraviolet_2012}). Studying FUV emission provides complementary constraints on accretion compared to more traditional diagnostics like H$\alpha$, because of the different temperature conditions that each probes. Since the maximum temperature reached depends on the shock velocity, studying FUV emission potentially provides different constraints on accretion flow itself compared to optical diagnostics like H$\alpha$.

In this paper, we analyse recent UV observations of the hot FUV atomic emission lines and warm molecular H$_2$ emission which occurs in the UV, together with nearly simultaneous optical data, of the brown dwarf 2MASS J08440915-7833457, hereafter J0844.  J0844 is a young, late M-type brown dwarf in $\eta$ Cha with evidence of accretion: it has an accretion disc based on infrared colour excess (\citealt{megeath_spitzerirac_nodate}) as well as excess continuum flux in its blue optical spectrum (\citealt{rugel_x-shooter_2018}).  

We compare our results to those from 2MASS J12073346-3932539 (hereafter J1207; \citealt{france_metal_2010}) and 2MASS J04141188+2811535 (hereafter J0414; \citealt{yang_far-ultraviolet_2012}). These objects are both accreting brown dwarfs with masses of 24 $M_\mathrm{jup}$ \citep{Riaz2007} and 65 $M_\mathrm{jup}$ \citep{herczeghillenbrand2008} respectively, and have been studied in terms of their FUV emission, though J0414 only has low-resolution observations. J1207 is particularly interesting in that, despite being an accreting source, it surprisingly showed C \textsc{\footnotesize{IV}} emission below the value expected from magnetic activity alone as defined by \cite{jk_2000}. \cite{france_metal_2010} note that J1207 has a weak magnetic field \citep{reiners2009} with likely no transition region activity due to the non-detection of Mg and Si species and conclude that the C \textsc{\footnotesize{IV}} emission seen is caused almost entirely by accretion. \cite{yang_far-ultraviolet_2012} measure the C \textsc{\footnotesize{IV}} of J0414 from low-resolution spectra, which we will use in the following sections. J0414 has a relatively high mass accretion rate for such a low-mass object (2.0 $\times$ $10^{-9}$ $M_\odot$ yr$^{-1}$; \citealt{herczeghillenbrand2008}). 

The paper is laid out as follows. Section \ref{subsec:fuv} contains an analysis of the FUV atomic lines, followed by an analysis of the H$_2$ emission in Section \ref{subsec:H2}, where we calculate inner and average emission radii. In Section \ref{subsec:optical}, we apply an accretion model to fit the nearly simultaneous optical continuum data in order to estimate an accretion rate. We then discuss the observed differences between the two brown dwarfs and other very low-mass stars, and speculate as to what causes these differences. We examine whether previous empirical relationships for low-mass stars can be extended to the brown dwarf regime, and what this may suggest in terms of accretion onto giant planets.

\begin{table*}[hbt!]
\caption{\textit{HST} observing log for J0844.}  
\centering
\begin{tabular}{c c c c c c c}
    \hline \hline
    Date & Start Time & Instrument & Mode & Wavelength Coverage & \textbf{Resolution} & $t_\mathrm{exp}$  \\
    & & & & \textbf{(\AA)} & ($\lambda/\Delta\lambda$) & (s) \\
    \hline
    2021-06-10 & 19:08:00 & COS & G130M & 1,132--1,436 & \textbf{$\sim$15,000} & 1950 \\ 
    2021-06-10 & 20:35:15 & COS & G160M & 1,417--1,792 & $\sim$15,000 & 4798 \\
    2021-06-10 & 23:52:36 & STIS & G230L & 1,570--3,180 & $\sim$1,000 & 1195 \\
    2021-06-11 & 01:21:56 & STIS & G430L & 2,900--5,700 & $\sim$1,000 & 2250 \\
    2021-06-11 & 00:20:56 & STIS & G750L & 5,240--10,270 & $\sim$1,000 & 260 \\
    \hline
\end{tabular}  
\label{tab:obs}
\end{table*}

\begin{figure}
    \begin{center}
    \includegraphics[width=\linewidth]{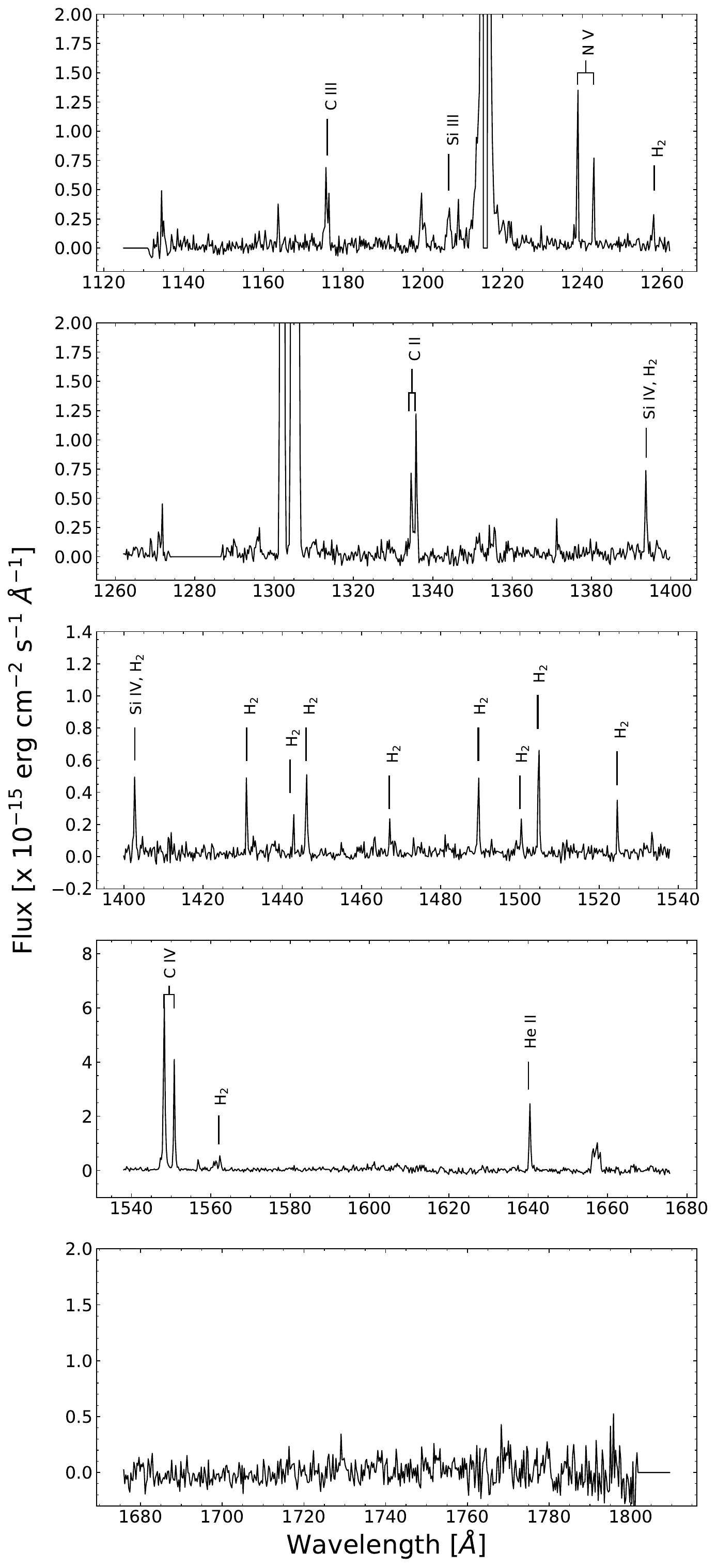}
    \end{center}
    \caption{The ultraviolet spectrum of J0844. The hot FUV lines are indicated, as well as the prominent molecular hydrogen lines. The prominent emission lines at $\lambda 1216$ \AA\ and $\lambda 1302$ \AA\ are the Ly$\alpha$ and O \textsc{\footnotesize{I}} airglow lines respectively.}
    \label{fig:uvspec}
\end{figure}

\section{Observations} \label{sec:obs}

J0844 (RA = 08$^\mathrm{h}$ 44$^\mathrm{m}$ 8.61$^\mathrm{s}$, DEC = --78$^\circ$ 33' 45.25'') was observed in June 2021 using the \textit{Hubble Space Telescope} (\textit{HST}) Cosmic Origins Spectrograph (COS) in TIMETAG mode and also with the Space Telescope Imaging Spectrometer (STIS). These data form part of the Hubble UV Legacy Library of Young Stars as Essential Standards (ULLYSES) Director's Discretionary Time (DDT) Program (\citealt{Roman-Duval_2020, romanduval2025}). This program was aimed at uniformly sampling very low-mass stars ($0.05$ $M_\odot < M_* < 0.5$ $M_\odot$) at medium resolution in the FUV, and at low resolution in the NUV-optical-IR while balancing the need to maintain reasonable exposure times. J0844 was the only brown dwarf included in the sample.

The data were reduced using the custom ULLYSES pipeline developed at the Space Telescope Science Institute (STScI). J0844 was observed in the FUV at medium resolution using COS/G130M and COS/G160M, and at low resolution using STIS/G230L, STIS/G430L and STIS/G750L in the NUV-optical-IR (see Table \ref{tab:obs} for an observing log). More information on the wavelength ranges, resolution, and signal-to-noise (S/N) ratios of these modes, as well as a description of the High Level Science Products (HLSPs) created by the ULLYSES team, can be found on the ULLYSES website\footnote{\url{https://ullyses.stsci.edu/index.html}}. 

J0844 is a 52 $M_\mathrm{jup}$ accreting brown dwarf (\citealt{rugel_x-shooter_2018}) located about $100$ pc away in the young $\eta$ Cha association, which is estimated to be about 8 million years old (\citealt{herczeg_empirical_2015}). Figure \ref{fig:uvspec} shows the UV spectrum of J0844, where the hot atomic FUV lines are indicated, as well as the prominent H$_2$ lines, which are pumped by Ly$\alpha$ photons generated near the accretion shock, as in TW Hya (e.g., \citealt{herczeg_farultraviolet_2002}). The prominent emission lines at $\lambda 1216$ \AA\ and $\lambda$1302 \AA\ are the Ly$\alpha$ and O \textsc{\footnotesize{I}} airglow lines respectively\footnote{\url{https://hst-docs.stsci.edu/cosihb /chapter-7-exposure-time-calculator-etc/7-4-detector-and-sky-backgrounds}}. We note that Figure \ref{fig:uvspec} has been binned to $\sim 0.7$ $\AA$, while the plots of various emission lines in the following sections have been binned to 5 km s$^{-1}$ for clarity.

\section{Analysis} \label{sec:results}

\subsection{Stellar Parameters} \label{subsec:params}

The stellar parameters of J0844 from \cite{rugel_x-shooter_2018} are listed in Table \ref{tab:params}. They used a distance to J0844 of 94 pc based on the distance to the $\eta$ Cha association. We use a distance of 99.6 pc based on the Gaia parallax for this object. As such, we adjust the parameters in Table \ref{tab:params} to account for this small difference in distance. We do not adjust the mass, as this object falls on the PMS Hayashi tracks (\citealt{baraffe2015}) meaning its mass is relatively constant for a small change in luminosity.

\cite{rugel_x-shooter_2018} find a radius of 0.36 $R_\odot$, which becomes 0.38 $R_\odot$ after adjusting for the small difference in distance. We find evidence that this radius may be somewhat too small, and we revise it as follows. The 2MASS $J$-band magnitude of this object is 12.505 (\citealt{brandeker2006}). Using equations from Section 4.6 of \cite{pecaut2013}, we estimate the luminosity using this $J$-band magnitude and the bolometric correction (BC) for the $J$-band in Table 6 of \cite{pecaut2013}. Unfortunately, these BCs are only given up to spectral type M5.0. If we extrapolate to later spectral types, taking into account how the BCs steepen towards M6.0, we get BC$_J$ = 2.139 which gives $R_* = 0.427$ $R_\odot$. Additionally, we estimate the stellar radius using theoretical spectra and scaling them by $R_*^2/d_*^2$ to match the observed spectrum, with $d_* = 99.6$ pc. We use BT-Settl (\citealt{allard2007, allard2011, allard2012, allard2013}), BT-NextGen (\citealt{allard2011, allard2012, husser2013}), and AMES-Cond (\citealt{allard2001, baraffe2003}) model spectra. For the observational data, we use an X-Shooter spectrum of J0844 taken as part of the PENELLOPE program \citep[see][]{manara2021} and  available from \url{https://zenodo.org/records/10024001}. There are two epochs of data; we use the first epoch which was obtained on 27 April 2021. The second epoch of data, taken on 1 May 2021, has very similar flux levels in the wavelength ranges we are concerned with, so we use only the first epoch. We fit for the radius in three primarily continuum regions in the red optical and near infrared:  $9,680-9,710$ \AA; $11,000-12,000$ \AA; and $15,000-17,000$ \AA. The models are defined by a specific $T_\mathrm{eff}$ and $\log g$. We use \cite{rugel_x-shooter_2018}'s parameters to find a starting point for $\log g = 3.99$. 
We use an iterative process to interpolate between the models in the grid (defined every 100 K and at log$g = 3.5$ and $4.0$) to match the observed spectra and find the radius at 2860 K.  We find $R_* = 0.425, 0.437, 0.444$ $R_\odot$ (the average across all three wavelength regions) for the BT-Settle, BT-NextGen, and AMES-Cond models respectively, finding an average of $R_* = 0.436$ $R_\odot$. If we average this value with the value we obtain using the BC, we get $R_* = 0.43$ $R_\odot$, which we quote in Table \ref{tab:params} and use in subsequent sections.

\begin{table}[hbt!]
\caption{Stellar parameters for J0844}
    \begin{center}
    \begin{tabular}{lcc}
         \hline \hline 
         Parameter & \cite{rugel_x-shooter_2018} value & Adopted value  \\
         \hline
         $M_*$ ($M_\odot$) & 0.052 & 0.052 \\
         $R_*$ ($R_\odot$) & 0.36 & 0.43 \\
         $L_*$ ($L_\odot$) & 0.008 & 0.009 \\
         SpT & M6.0 & M6.0 \\
         $A_\mathrm{V}$ (mag) & 0.0 & 0.0 \\
         $T_{\mathrm{eff}}$ (K) & 2860 & 2860 \\
         $d$ (pc) & 94.0 & 99.6 \\
         \hline
    \end{tabular}
    \end{center}
    \tablecomments{The adopted values differ from the \cite{rugel_x-shooter_2018} values due to the assumed difference in distance, with the exception of the radius (see text for further details). }
    \label{tab:params}
\end{table}


The amount of FUV flux produced through magnetic activity is proportional to the inverse Rossby number, sometimes referred to as the dynamo number, which is the convective turnover time divided by the rotation period (\citealt{hartmann1987}). Therefore, we need to determine rotation periods of J0844 and its comparison objects. In addition to J1207 and J0414 discussed in Section \ref{sec:intro}, we choose to compare J0844 to the magnetically active, late M dwarfs VB 8, VB 10, and LHS 2065 and the magnetically active, mid M dwarf EV Lac which were observed in the FUV by \cite{hawley_transition_2003} and \cite{osten_radio_2006} respectively. 

\begin{figure*}
    \centering
    \includegraphics[width=\linewidth]{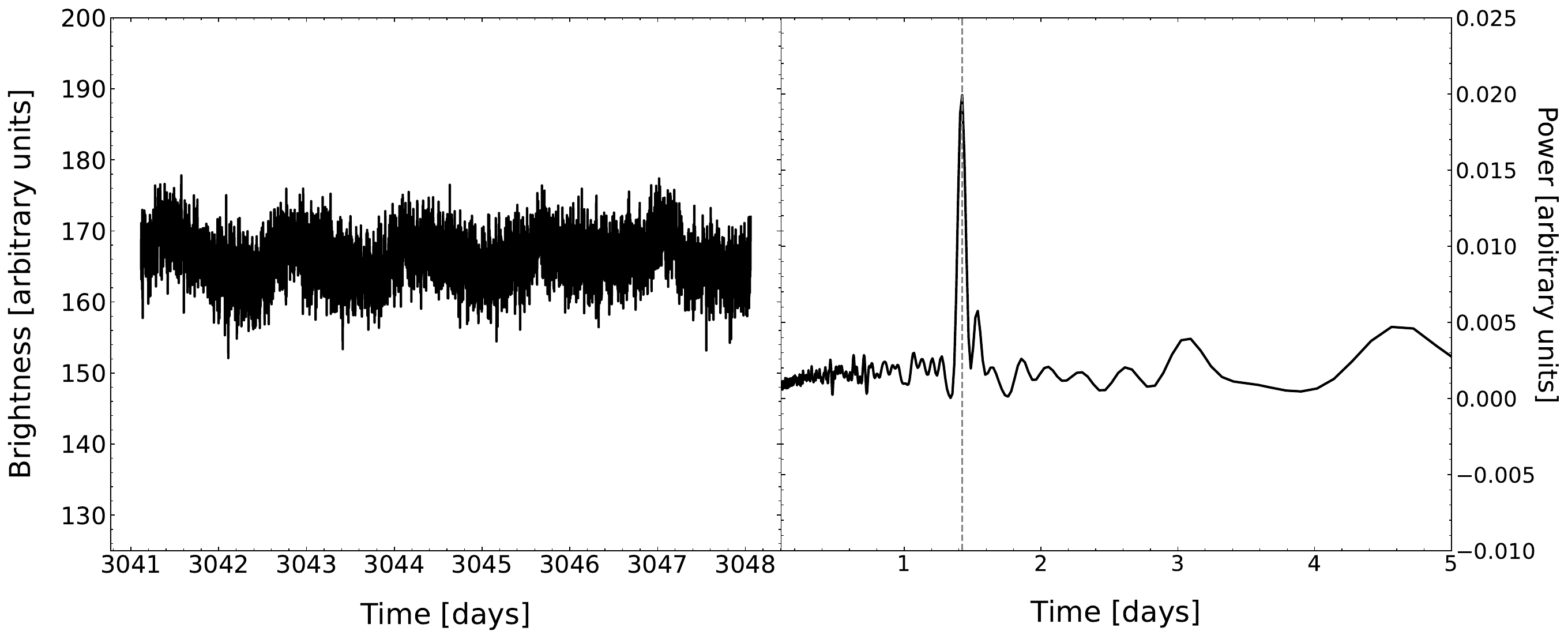}
    \caption{Left panel: Portion of the lightcurve of J0844 as observed with TESS, showing cyclic variations attributed to rotation. Right panel: Power spectrum obtained through a LombScargle periodogram analysis (\citealt{2018ascl.soft12013L,2018AJ....156..123A}) of the left panel. The grey dashed line marks the power spectrum peak, which corresponds to a period of 1.42 days (Table \ref{tesslog} and \ref{tab:periods}.)}
    \label{fig:lc}
\end{figure*}

Rotation periods are commonly determined by performing a periodogram analysis on an object's lightcurve. We use lightcurves from TESS (see Table \ref{tesslog} for observing log and recovered periods), except for VB 8, which has not been observed by TESS. Each object has multiple sector observations; however, we simply choose one sector with a clean observation for our analysis. We use \textsc{astropy}'s LombScargle function (\citealt{2013A&A...558A..33A}) for the periodogram analysis. A portion of J0844's TESS lightcurve and its corresponding power spectrum are shown in Figure \ref{fig:lc}. The lightcurves of VB 10, LHS 2065, and EV Lac show flares, so these were removed from the data by interpolating across them before doing the periodogram analysis. 

TESS has a large pixel size of about 21 arcseconds, so in order to ensure no contamination from nearby sources, a field of view of $\sim$25 arcseconds was investigated around each target using SIMBAD (\citealt{simbad}) DSS images, since these are in the optical band which matches that of TESS. J0844, J0414, and LHS 2065 appear single. J1207 is a binary system consisting of M8 and L3 brown dwarfs separated by an angular distance of $\sim$0.78 arcseconds (\citealt{luhman_jwstnirspec_2023}), meaning both objects will fall well within a single TESS pixel. However, the object we are interested in is the M8 brown dwarf, which is more than 400 times brighter than the L3 brown dwarf in the J-band \citep{luhman_jwstnirspec_2023}. The TESS bandpass is shortwards of the J-band, and we expect the flux ratio to be even greater at these shorter wavelengths. Therefore, we assume the recovered period is for the M8 object. We note that our derived period agrees with \citet{Koen2008} who found evidence for rotational modulation produced by accretion hot spots in J1207 which also imply a period of about a day. Both VB 10 and EV Lac appear to be blended with another object, and so their lightcurves may be contaminated.

As a comparison to the recovered TESS periods, we also calculate an upper limit on the period for each object using literature $v\sin i$ measurements when available. We take a weighted average if the object has multiple $v\sin i$ estimates (Table \ref{tab:my_label}), excluding values which are very different to the other measurements. The radius we use for each object is $R_* = 0.12$ $R_\odot$ for VB 8 and VB 10 (\citealt{pineda_m-dwarf_2021}), $R_* = 0.11$ $R_\odot$ for LHS 2065 (\citealt{pineda_m-dwarf_2021}), and $R_* = 0.3$ $R_\odot$ for EV Lac (\citealt{osten_radio_2006}). 


The TESS, $v \sin i$, and final adopted periods are shown in Table \ref{tab:periods}. We use the TESS period for J0844, J1207, and J0414, since these objects do not have literature $v \sin i$ estimates. VB 8 has not been observed by TESS, so we use the calculated $v \sin i$ period. Since VB 10 appears to be blended with another object, we also use the calculated $v \sin i$ period. While EV Lac is also a blended source, the recovered TESS period is consistent with previous measurements (e.g., \citealt{jk_2000}) and so we use the TESS period. 

We note that it does not matter too much which period we use, since the parameter we are concerned with is the dynamo number (convective turnover time, $\tau_\mathrm{c}$, divided by rotation period, $P_\mathrm{rot}$), and all of these objects remain in the saturated regime of the rotation--activity relation discussed below whether we use the TESS or $v \sin i$ derived period. For the convective turnover time, we use 180 days for all objects based on Figure 2 in \cite{Gilliland_1986}. This is also consistent with later work on convective turnover times (see Figure 7 in \citealt{wright2011}), which estimates convective turnover times to be greater than $\sim$150 days at these low masses. 


\begin{table}[hbt!]
\caption{TESS observing log.}
\begin{center}
    \begin{tabular}{l c c c l }
        \hline \hline
         Target & Start Date & End Date & Sector & Period (days) \\
         \hline 
         J0844 & 2023-04-06 & 2023-05-04 & 64 & 1.42 $\pm$ 0.03 \\
         J1207 & 2021-04-02 & 2021-04-28 & 37 & 0.98 $\pm$ 0.04 \\
         J0414 & 2021-09-16 & 2021-10-12 & 43 & 3.00 $\pm$ 0.59 \\
         VB 10 & 2022-07-09 & 2022-08-04 & 54 & 3.91 $\pm$ 0.12  \\
         LHS 2065 & 2019-02-02 & 2019-02-27 & 8 & 0.46 $\pm$ 0.05 \\
         EV Lac & 2019-09-11 & 2019-10-06 & 16 & 4.35 $\pm$ 0.40 \\
         \hline
    \end{tabular}
    \end{center}
    \tablecomments{Each object had multiple sector observations; however, we choose to only analyse one sector for each object, choosing a sector with a clean observation. The uncertainty we quote is the half-width at half-maximum of a Gaussian fit to the peak of the power spectrum, an example of which is shown in the right panel of Figure \ref{fig:lc}.}
    \label{tesslog}
\end{table}

\begin{table}[hbt!]
    \caption{$v \sin i$ measurements}
    \begin{center}
    \begin{tabular}{l l l}
        \hline \hline
         Target & $v\sin i$ (km s$^{-1}$) & Reference  \\
         \hline
         VB 8 & 3.67 $\pm$ 1.07 & \cite{han_magnetic_2023} \\
         & 6.07 $\pm$ 0.68 & \cite{han_magnetic_2023} \\
         & $^\dagger$10.1 $\pm$ 0.8 & \cite{fouque_spirou_2018} \\
         & 5.4 $\pm$ 1.5 & \cite{reiners_carmenes_2018} \\
         & 5.38 $\pm$ 0.54 & Weighted average \\
         \hline
         VB 10 & 5.86 $\pm$ 0.78 & \cite{han_magnetic_2023}\\
         & 5.47 $\pm$ 1.05 & \cite{han_magnetic_2023} \\
         & 5.3 $\pm$ 0.9 & \cite{fouque_spirou_2018} \\
         & 2.7 $\pm$ 2.2 & \cite{reiners_carmenes_2018} \\
         & 5.43 $\pm$ 0.50 & Weighted average \\
         \hline
         LHS 2065 & 9.3 $\pm$ 2.8 & \cite{reiners_carmenes_2018} \\
         \hline
         EV Lac & 6.9 $\pm$ 0.8 & \cite{glebocki2005} \\
         & $^\dagger$10.9 $\pm$ 1.9 & \cite{glebocki2005} \\
         & 4.89 $\pm$ 0.79 & \cite{han_magnetic_2023} \\
         & 3.18 $\pm$ 1.18 & \cite{han_magnetic_2023} \\
         & 5.9 $\pm$ 0.7 & \cite{fouque_spirou_2018} \\
         & 3.5 $\pm$ 1.5 & \cite{reiners_carmenes_2018} \\
         & 4.5 $\pm$ 0.5 & \cite{jk1996} \\
         & 5.06 $\pm$ 0.31 & Weighted average \\
         \hline
    \end{tabular}
    \end{center}
    \label{tab:my_label}
    \tablecomments{$^\dagger$These value were excluded from the weighted averages. See text for further details.}
\end{table}

\begin{table}[hbt!]
\caption{Estimated rotation periods}
    \begin{center}
    \begin{tabular}{l c c c}
        \hline \hline
         Target & TESS Period & $v\sin i$ Period & Adopted Period \\
         & (days) & (days) & (days)  \\
        \hline
        J0844 & 1.42 $\pm$ 0.03 & -- & 1.42 \\
        J1207  & 0.98 $\pm$ 0.04 & -- & 0.98 \\
        J0414 & 3.00 $\pm$ 0.59 & -- & 3.00 \\
        VB 8 & -- & $<$ 1.13 + 0.11 & 1.13 \\
        VB 10 & 3.91 $\pm$ 0.30 & $<$ 1.12 + 0.10 & 1.12 \\
        LHS 2065 & 0.46 $\pm$ 0.05 & $<$ 0.60 + 0.20 & 0.46 \\
        EV Lac & 4.35 $\pm$ 0.40 &  $<$ 3.00 + 0.20 & 4.35 \\
        \hline
    \end{tabular}
    \end{center}
    \tablecomments{The uncertainty we quote for the TESS period the half-width at half-maximum of a Gaussian fit to the peak of the power spectrum shown in the left panel of Figure \ref{fig:lc}. The period based on $v \sin i$ is an upper limit, as the equatorial velocity could be higher. We thus show these values with a 1$\sigma$ (upper) errorbar. In the last column, we show the final adopted period (either from TESS or from $v\sin i$) that is used in following sections.}
    \label{tab:periods}
\end{table}

\subsection{The Far-Ultraviolet (FUV) Spectrum}
\label{subsec:fuv}

\begin{table*}[hbt!]
\caption{FUV atomic line characteristics}
    \begin{center}
    \begin{tabular}{llcccccc}
         \hline \hline 
         ID & $\lambda_\mathrm{rest}$ & $v$ & $\sigma(v)$ & FWHM & $\sigma$(FWHM) & Line Flux & $\sigma$(Line Flux) \\
         & (\AA) & (km s$^{-1}$) & (km s$^{-1}$) & (km s$^{-1}$) & (km s$^{-1}$) & ($\times 10^{-16}$ ergs cm$^{-2}$ s$^{-1}$) & ($\times 10^{-16}$ ergs cm$^{-2}$ s$^{-1}$) \\ 
         \hline
         C \textsc{\footnotesize{III}} & 1176.0 & -- & -- & -- & -- & 2.7 & 0.66 \\
         \hline
         Si \textsc{\footnotesize{III}} & 1206.5 & -- & -- & -- & -- & 1.62 & 0.55 \\
         \hline
         N \textsc{\footnotesize{V}} & 1238.8 & 20.45 & 1.35 & 48.09 & 4.48 & 5.76 & 0.63 \\
         & 1242.8 & 25.60 & 5.00 & 57.13 & 14.55 & 3.06 & 0.52 \\
         \hline
         C \textsc{\footnotesize{II}}  & 1334.53 & -- & -- & -- & -- & 1.67 & 0.55 \\
         \hline
         C \textsc{\footnotesize{II}}  & 1335.71 & -- &  -- & -- & -- & 3.84 & 0.58 \\
         \hline
         Si \textsc{\footnotesize{IV}} & 1393.8 & -- & --  & --  & -- & 1.71 & 0.93 \\
           & 1402.8 & -- & -- & -- & -- & 0.92 & 0.79 \\
           \hline
          C \textsc{\footnotesize{IV}} & 1548.2 NC & 28.73 & 3.08 & 35.44 & 23.19 & 34.61 & 1.08 \\
           & 1548.2 BC & 41.16 & 6.47 & 135.66 & 32.29 &  &  \\
         & 1550.8 NC & 23.21 & 4.14 & 25.17 & 12.28 & 15.96 & 0.75 \\
         & 1550.8 BC & 38.30 & 19.33 & 94.92 & 26.45 &  &  \\
         \hline
        He \textsc{\footnotesize{II}}  & 1640.4 & 30.83 & 5.54 & 67.47 & 9.24 & 8.31 & 0.79 \\
         \hline
    \end{tabular}
    \end{center}
    \tablecomments{Characteristics of the lines identified in Figure \ref{fig:uvspec}. Each line is assumed to be singly Gaussian in shape, except C \textsc{\footnotesize{IV}}, which is fit with a narrow component (NC) and a broad component (BC). See Section \ref{subsec:fuv} for details.}
    \label{tab:linefluxes}
\end{table*}

We analyse several of the stronger hot FUV atomic emission lines profiles labelled in Figure \ref{fig:uvspec}. Table \ref{tab:linefluxes} shows the derived line profile properties, and the measured line fluxes. 

These lines are produced at temperatures around $10^5$ K. This temperature is potentially reached both on the stellar surface due to transition region (magnetic) activity, as well as in the accretion flows at the accretion shock. The transition region emission should be relatively narrow, broadened thermally and by the rotation of the star. Accretion activity, however, could contribute emission at higher velocities, since the material is travelling much faster, and therefore could give rise to a broad emission component. Therefore, each emission line might be a superposition of two components, with one broader than the other.

\begin{figure*}
    \includegraphics[width=\linewidth]{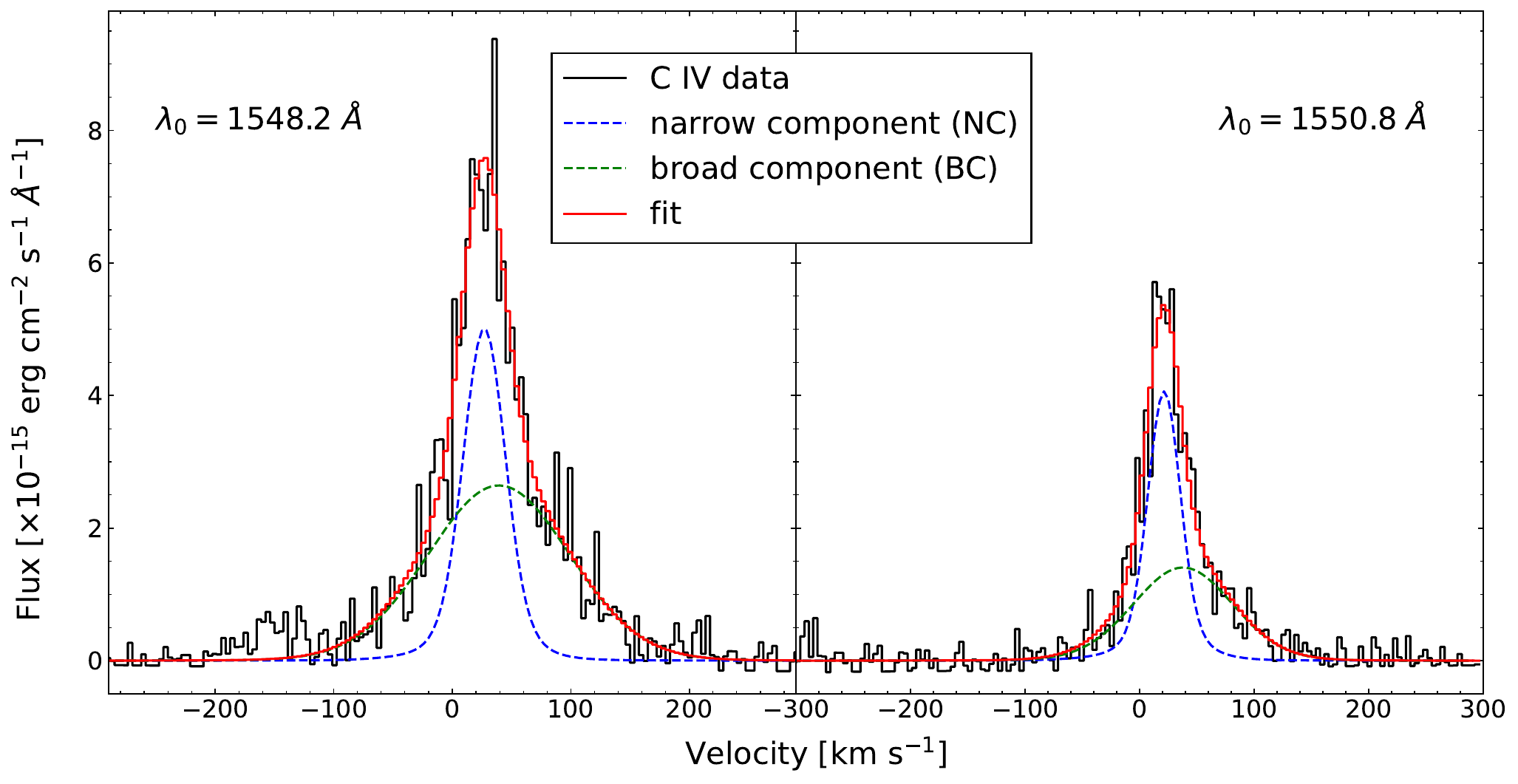}
    \caption{Fits to the C \textsc{\footnotesize{IV}} doublet. The left panel is the first line of the doublet, at $\lambda$1548.2 \AA. The right panel is the second line of the doublet, at $\lambda$1550.8 \AA. The zero velocity on each panel corresponds to these rest wavelengths. The green and blue curves are the two Gaussian components, and the red curve is the best fit (addition of components convolved with the LSF).}
    \label{fig:civfit}
\end{figure*}

We thus test whether each line is best fit by a single or double Gaussian: we fit each line profile with a double Gaussian and then a single Gaussian (incorporating the COS LSF at an appropriate tabulated wavelength) using Python's \textsc{scipy}, after subtracting a linear continuum. We estimate the continuum contribution by interpolating between two small regions on the red and blue sides of each line. The fit with the lower reduced $\chi^2$ value is adopted. This turns out to be a single Gaussian for all lines except C \textsc{\footnotesize{IV}}, which is better fit with broad and narrow Gaussian components (Figure \ref{fig:civfit}). The reason for this difference between C \textsc{\footnotesize{IV}} and all the other lines may be real; however, since C \textsc{\footnotesize{IV}} is the strongest line (Figure \ref{fig:uvspec}), it is also possible that the S/N of the other lines is simply too low to resolve individual broad and narrow components.

To test whether the S/N is playing a role, we use the double Gaussian fit obtained for C \textsc{\footnotesize{IV}} 1548.2 \AA\ and scale it to match the flux level of the stronger of the two N \textsc{\footnotesize{V}} lines at 1238.2 \AA. We then add noise at the observed noise level of the N \textsc{\footnotesize{V}} line, and refit the profile using both a double and single Gaussian, to see whether we indeed still recover that the double Gaussian is a better fit at the S/N of the N \textsc{\footnotesize{V}} line compared to the single Gaussian. We repeat this process 500 times. The double Gaussian is a better fit only about 50\% of the time, suggesting that S/N is possibly contributing to the difference in shape between the stronger C \textsc{\footnotesize{IV}} lines and the other weaker line species.

The fit parameter uncertainties are calculated using a simple Monte Carlo routine, in which random noise at the observed level is added to the fitted spectrum to create 1,000 iteration of the line profiles, and the standard deviation of each collection of fit parameters is used as the uncertainty for that parameter. While flux uncertainties in the data are often assumed to be normally distributed, the distribution of the uncertainties in these data is not well fitted by a normal distribution. To estimate the uncertainty distribution, we plot a histogram of the fit residuals and find that this distribution is better fit by a Lorentzian profile. We then sample randomly from this Lorentzian profile as a measure of the uncertainty at each point.  We then visually check the resulting histogram of modelled uncertainties to verify that it closely resembles that found in the original data. The S/N was too low to extract meaningful Gaussian fit parameters from the C \textsc{\footnotesize{III}}, Si \textsc{\footnotesize{III}} and C \textsc{\footnotesize{II}} lines, so we measure only the integrated flux as described below. 

The line flux for all lines is calculated by integrating across the observed line profile, after removing any H$_2$ lines present (e.g., in C \textsc{\footnotesize{IV}} $\lambda$1548.2 \AA\, left panel of Figure \ref{fig:civfit}) and subtracting a linear continuum. Again, we estimate the continuum contribution by interpolating between two small regions on the red and blue sides of each line. 
We choose to integrate the line profile itself rather than using the Gaussian fit parameters to calculate the line flux. This is because the lines may not be truly Gaussian in shape; however, calculating the line flux with the fit parameters agrees with that calculated by integrating the data itself within the uncertainties. 

\begin{figure*}

    \includegraphics[width=\linewidth]{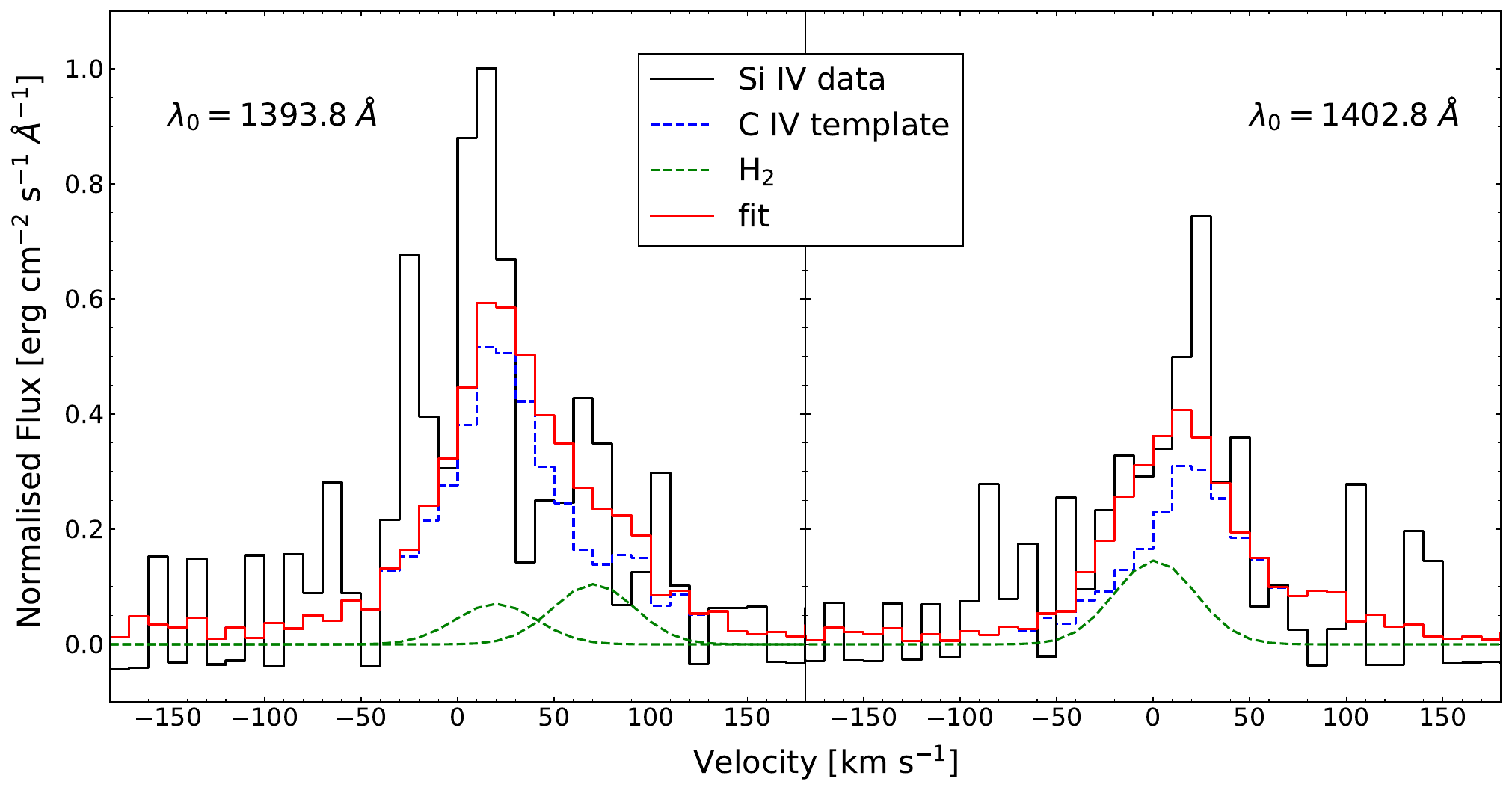}
    \caption{Fits to the Si \textsc{\footnotesize{IV}} doublet. The left panel is the first line of the doublet, at $\lambda$1393.8 \AA. The right panel is the second line of the doublet, at $\lambda$1402.8 \AA. The zero velocity on each panel corresponds to these rest wavelengths. The blue curve is the C \textsc{\footnotesize{IV}} template. The H$_2$ lines (green curves) lie at rest wavelengths of $\lambda$1393.73, 1393.96 \AA\ (left panel), and $\lambda$1402.65 \AA\ (right panel). These lines were assumed to be present based on the H$_2$ emission observed in TW Hya (\citealt{herczeg_farultraviolet_2002}). The red curve is the best fit. Given the relatively high noise levels in this region, these data have been binned to 10 km s$^{-1}$ instead of 5 km s$^{-1}$ as is the case for all other emission line plots.} 
    \label{fig:siivfit}
\end{figure*}

The Si \textsc{\footnotesize{IV}} line fluxes are estimates only, as the amount of Si \textsc{\footnotesize{IV}} present is not as obvious. The data are noisy, and a few relatively strong H$_2$ lines are expected to lie near the observed emission lines based on the H$_2$ observed in other CTTSs like TW Hya (\citealt{herczeg_farultraviolet_2002}). Conceivably, the strength of these H$_2$ lines near Si \textsc{\footnotesize{IV}} could be inferred from branching ratios in combination with measurements of stronger H$_2$ lines at different wavelengths but from the same progression. 
However, as we show in Section \ref{subsec:H2}, the strongest observed H$_2$ lines are from different progressions and therefore pumped by different parts of the Ly$\alpha$ line. These detections are already individually weak, so the lines needed to use the branching ratios are even weaker. Therefore, in order to estimate how much Si \textsc{\footnotesize{IV}} is present, we use the observed C \textsc{\footnotesize{IV}} $\lambda$1548.8 \AA\ line as a template for the shape of both lines of the Si \textsc{\footnotesize{IV}} doublet, and add Gaussians for the H$_2$ lines to create a composite spectrum. The ratio of the Si \textsc{\footnotesize{IV}} $\lambda 1393.8$ \AA\ line to the $\lambda 1402.8$ \AA\ is estimated to be 0.6, based on the ratio observed in EV Lac (\citealt{osten_radio_2006}), an active M dwarf with strong magnetic fields (\citealt{jk1996}). The ratio of the H$_2$ lines to one another is assumed to be the same as those found for TW Hya (\citealt{herczeg_farultraviolet_2002}). This leaves two free parameters: the overall strength of the Si \textsc{\footnotesize{IV}} lines, and the overall strength of the H$_2$ lines. The best fit to the data is then determined by varying the strengths of these Si \textsc{\footnotesize{IV}} and H$_2$ lines simultaneously, and using a $\chi^2$ minimisation. The final scaled strengths of the C \textsc{\footnotesize{IV}} templates (as proxies for the Si \textsc{\footnotesize{IV}}) are then used as an estimate of the amount of Si \textsc{\footnotesize{IV}} present (see Figure \ref{fig:siivfit}), and integrated in the same way as described for the other lines. The uncertainties are calculated in the same way as for the other lines using a Monte Carlo simulation. 

While none of the other lines in the progressions corresponding to the H$_2$ lines near Si \textsc{\footnotesize{IV}} are individually detected (and therefore make branching ratio calculations less reliable), we still integrate the flux in the regions where these lines are expected to be in order to predict the flux in the $\lambda1393.73$, $\lambda 1393.96$, and $\lambda 1402.65$ lines which are the ones that overlap in wavelength with the Si \textsc{\footnotesize{IV}} lines. The $\lambda1393.73$ line is pumped by $\lambda 1217.205$, while the other two are pumped by $\lambda 1217.643$. We consider lines pumped by these two wavelengths as identified in \cite{herczeg_farultraviolet_2002}, and define a consistent velocity range over which to integrate. Using each estimated line flux pumped by $\lambda 1217.643$, we predict the strength of the $\lambda 1402.65$ and $\lambda 1393.96$ lines and take a weighted mean of the different predictions as the estimate of the expected line strength. We repeat this process for the lines pumped by $\lambda 1217.205$. For the three lines we are trying to estimate, the predicted flux levels are swamped by the uncertainty in the predicted flux, reflecting just how weak each of the individual H$_2$ lines are from these progressions.  However, given the final uncertainty, these estimates are consistent with what we find from the fits in Figure \ref{fig:siivfit}. We further explore the origin of J0844's H$_2$ emission in Section \ref{subsec:H2}.



\subsection{Molecular Hydrogen} \label{subsec:H2}

Molecular hydrogen emission is ubiquitous in systems with circumstellar discs (\citealt{herczeg_farultraviolet_2002, france_radial_2023}). Ly$\alpha$ photons photoexcite H$_2$ molecules to higher electronic states, which then emit in the UV when they de-excite (\citealt{herczeg_farultraviolet_2002}). In order for H$_2$ to absorb the Ly$\alpha$ photons, the molecule must be excited above the ground state. This happens thermally in gas with temperatures of around 1000 -- 4000 K (\citealt{herczeg2004, Flagg2022}), which is achieved both in the brown dwarf photosphere and in the inner disc. This means that any observed H$_2$ emission could originate from both the star and the disc. 

Studying molecular emission close to the star can give information on the inner disc structure, including its truncation radius, which may indicate whether magnetic fields from the central object truncate the inner disc. \cite{france_radial_2023} do not report any H$_2$ in the J0844 system, although we see a few individual H$_2$ lines detected (Figure \ref{fig:uvspec}) and we list their transitions and measured line fluxes in Table \ref{tab:h2}. However, the S/N at each individual line is low. To improve the S/N, we follow \cite{france_metal_2010} and co-add the five strongest H$_2$ lines from Table \ref{tab:h2}. These are four lines from the [1,4] progression and one line from the [1,7] progression. While different progressions are produced by different parts of the Ly$\alpha$ line and therefore can give insights into the strength and shape of the Ly$\alpha$ emission, this does not change where the H$_2$ emission arises from in the star and/or disc, which is what we are most interested in. Therefore, we combine [1,4] and [1,7] progression lines to achieve the best S/N for the coadded line profile. Though the $\lambda 1547.34$ \AA\ line is the one of the stronger H$_2$ lines, we do not include it in Table \ref{tab:h2} or in the coaddition because it lies right next to the C \textsc{\footnotesize{IV}} $\lambda 1549$ \AA\ line, and so may suffer contamination from this strong line. We also note that the H$_2$ detections come from only two progressions, pumped by $\lambda 1216.07$ \AA\ and $\lambda1215.73$ \AA, which are both fairly close to the center of the Ly$\alpha$ line where the Ly$\alpha$ emission is expected to be the strongest \citep{herczeg2004}. The lack of H$_2$ line detections from progressions pumped by wavelengths further from the Ly$\alpha$ line core suggests that the Ly$\alpha$ emission line may be relatively narrow compared to typical CTTSs \citep[see also][]{Flagg2021, Flagg2022}. We subtract the continuum from our coadded H$_2$ profile by fitting a line to the blue and red sides of the line profile, in the same way as when calculating the line flux in Section \ref{subsec:fuv}.

\subsubsection{Double Gaussian Fit} \label{subsubsec:doublegauss}

\begin{figure}[hbt!]
    \includegraphics[width=\linewidth]{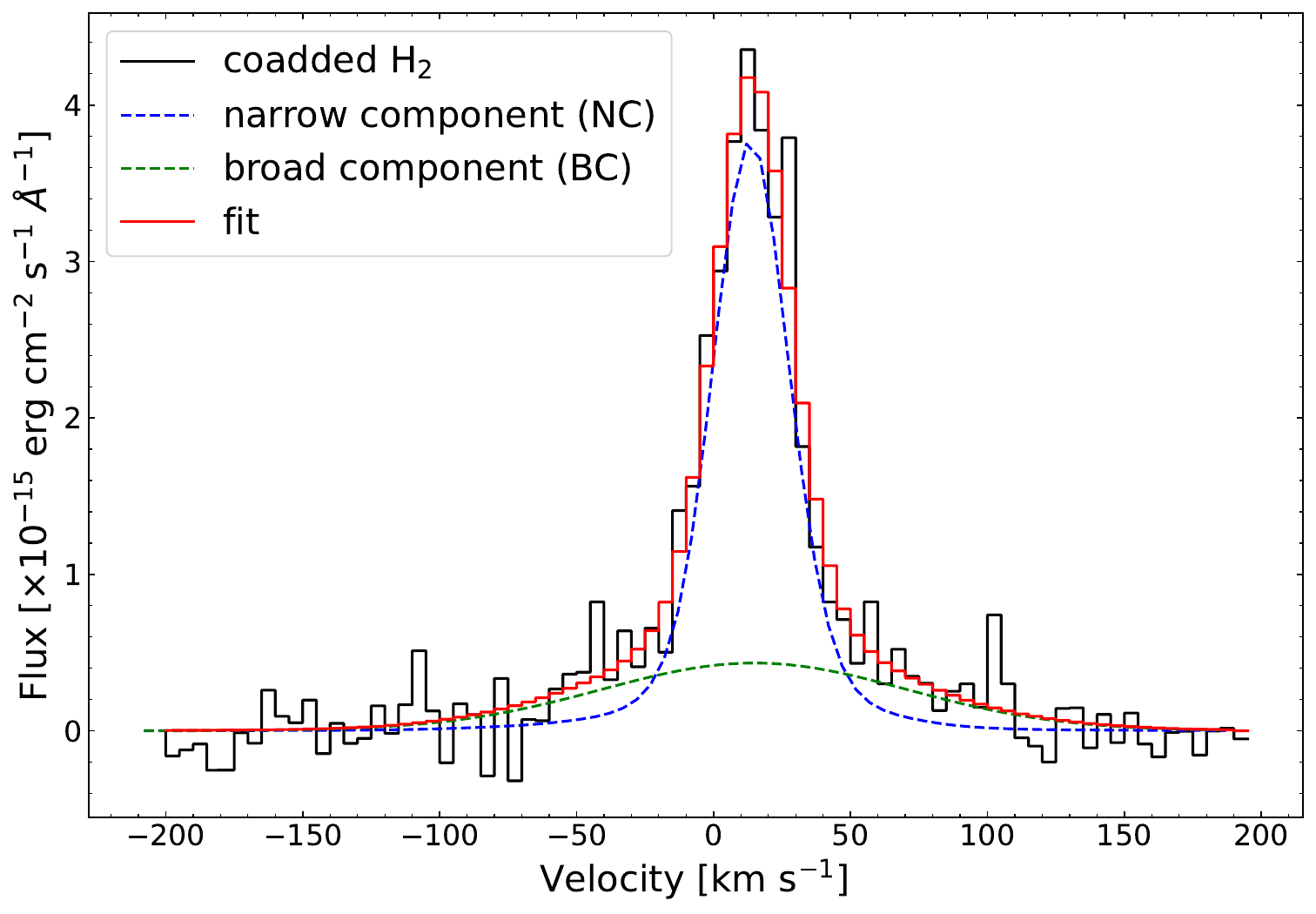}
    \caption{Double Gaussian fit to the coadded H$_2$ line profile. The blue curve is the narrow component, and the green curve is the broad component. The red curve is the best fit (addition of components convolved with the LSF). The narrow component has $v = 12.8$ $\pm$ 6.0 km s$^{-1}$ and FWHM = 24.2 $\pm$ 9.6 km s$^{-1}$. The broad component has $v = 14.0$ $\pm$ 10.7 km s$^{-1}$ and FWHM = 126.5 $\pm$ 23.5 km s$^{-1}$. The fit has $\chi^2 = 29.6$, and $\chi^2_\mathrm{red} = 0.4$.}
    \label{fig:h2doublegauss}
\end{figure}

We fit the observed (continuum-subtracted) coadded line profile in two ways. Firstly, we use a double Gaussian fit as shown in Figure \ref{fig:h2doublegauss}, taking into account the COS LSF as described in Section \ref{subsec:fuv}. This is often how H$_2$ profiles have been analysed (\citealt{france_hubble_2012,gangi_penellope_2023}), under the general assumption that the narrower component is stellar in origin, and the broader component originates from the disc. The width of the broad component (accounting for inclination) can then used to estimate the velocity at which the emission originates, which can be translated into an emission radius by assuming that Keplerian motion is dominating the line broadening. We find the width of the two components are 24.2 $\pm$ 9.6 km s$^{-1}$ for the narrow component and 126.5 $\pm$ 23.5 km s$^{-1}$ for the broad component.

\begin{table}[hbt!]
    \caption{H$_2$ lines observed in both J1207 and J0844.}
    \begin{center}
    \begin{tabular}{l c c c c c}
    \hline \hline
         Transition & [$v'$, $J'$] & $\lambda_\mathrm{rest}$ & $\lambda_\mathrm{pump}$ & Line Flux ($\sigma$)  \\
         & & (\AA) & (\AA) \\
         \hline
         R(3) 1--3 & [1,4] & 1257.83 & 1216.07 & 0.67 (0.43) \\
         $^\dagger$R(3) 1--6 & [1,4] & 1431.01 & 1216.07 & 1.58 (0.43) \\ 
         $^\dagger$R(3) 1--7 & [1,4] & 1489.57 &1216.07 & 1.62 (0.28) \\
         $^\dagger$P(5) 1--6 & [1,4] & 1446.12 &1216.07 & 2.00 (0.28) \\
         $^\dagger$P(5) 1--7 & [1,4] & 1504.76 &1216.07 & 3.11 (0.37) \\ 
         P(5) 1--8 & [1,4] & 1562.39 &1216.07 & 1.07 (0.29)  \\
         
         R(6) 1--6 & [1,7] & 1442.87 &1215.73 & 0.73 (0.24)  \\
         R(6) 1--7 & [1,7] & 1500.45 &1215.73 & 0.86 (0.31) \\
         P(8) 1--6 & [1,7] & 1467.08 &1215.73 &  0.49 (0.33) \\
         
         $^\dagger$P(8) 1--7 & [1,7] & 1524.65 &1215.73 &  1.73 (0.36) \\ 
        \hline
    \end{tabular}
    \end{center}
    \tablecomments{The upper electronic state is characterised by $[v', J']$, and the lower by $[v'', J'']$; the number in parentheses after the $R$ and $P$ designations is $J''$. $R$ denotes $\Delta J = -1$ and $P$ denotes $\Delta J = +1$ where $\Delta J = J' - J''$. The hyphenated numbers are the vibrational states $v'$--$ v''$. The fluxes are in units of $10^{-16}$ erg cm$^{-2}$ s$^{-1}$. \\ $^\dagger$These lines were used in the coaddition to create Figure \ref{fig:h2doublegauss} and \ref{fig:h2disc}.}
    \label{tab:h2}
\end{table}

\subsubsection{Stellar and Disc Component Fit} \label{subsubsec:disc}

\begin{figure}[hbt!]
    \centering
    \includegraphics[width=\linewidth]{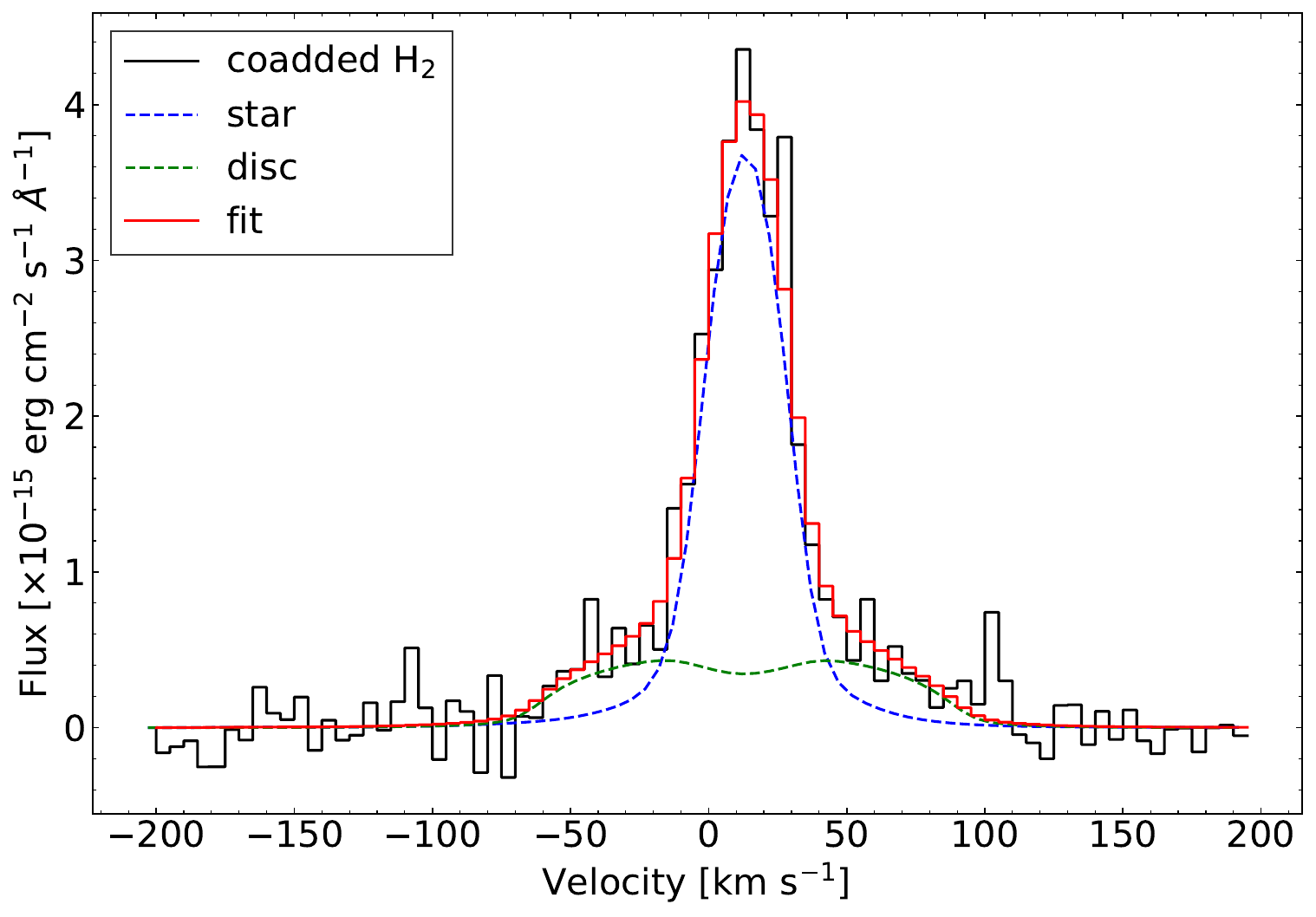}
    \caption{The blue curve is the stellar component, and the green is the disc component. The red curve is the best fit (addition of components convolved with the LSF). We set $i_\mathrm{disc} = 80^\circ$, with RV = $12.30$ km s$^{-1}$ and $v\sin i = 18.67$ km s$^{-1}$, and recover an inner radius of $R_\mathrm{in} = 3.58$ $R_*$ (see text for details). The fit has $\chi^2 = 27.87$, and $\chi^2_\mathrm{red} = 0.35$.}
    \label{fig:h2disc}
\end{figure}

We assumed in the previous calculations that the narrow Gaussian component of the H$_2$ emission is coming from the star, and the broad Gaussian component is coming from the disc. Since neither the star nor the disc profile are intrinsically Gaussian, we explore a more physical model representing a true stellar component and a full disc component. 

For the stellar emission, we start with a Gaussian as the intrinsic line shape at a thermal temperature of 2500 K, which corresponds to a line width in velocity space of about 5 km s$^{-1}$. We choose this temperature and its corresponding velocity because the temperatures needed for H$_2$ emission are around 1000 -- 4000 K as mentioned previously (\citealt{herczeg2004,Flagg2022}). We also examine a model photosphere with J0844's $T_\mathrm{eff}$ using the BT-NextGen models and find that 2500 K is indeed reached within the stellar atmosphere, and thus is a reasonable guess. Then, we broaden the profile to account for rotation (letting $v \sin i$ be a free parameter), using the rotational broadening code from \cite{carvalho_simple_2023}, and use this as our estimation for the stellar emission. 


The disc model calculates the emission (assumed to be locally Gaussian, whose width is specified by a turbulent velocity, $v_\mathrm{turb}$) at each unit area of the disc and integrates over the disc to create the full emission profile. The disc velocity at the stellar surface can be calculated according to $v_\mathrm{surf} = \sqrt{G M_*/R*}$ and the orbital velocity is assumed to fall off with distance as a Keplerian. We set $v_\mathrm{turb}$ = 1 km s$^{-1}$ everywhere in the disc. This is significantly smaller than the orbital velocity and assumes the line broadening we see is dominated by Keplerian motion.

For emission from systems with circumstellar discs, the relationship between intensity and radius is usually assumed to be a power, such as in \cite{salyk_co_2011} where they fit infrared CO disc emission. However, as they discuss, the exact power is not well-constrained, though they find that $p \leq -1.5$ generally ensures a robust fit for $R_\mathrm{in}$. Molecular hydrogen disc emission differs from CO in that it is assumed to be caused entirely by fluorescence (i.e., no thermal contribution). In general, for a flat disc illuminated by a central star only, the intensity at the disc is calculated using 

\begin{equation}
    I(r) = I_0 \Biggl[\arcsin{\Biggl(\frac{R_*}{r}\Biggr)} - \frac{R_*}{r}\Biggl( 1 - \frac{R_*^2}{r^2}\Biggr)^{1/2} \Biggr]
    \label{ieq}
\end{equation}
from \cite{staracc}, and we assume that the H$_2$ emission follows the same law since it is produced through fluorescence. In the limit where $r \gg R_*$, this equation reduces to a power law with $p = -3$: a steeper falloff than what \cite{salyk_co_2011} used; however, we are assuming no thermal contribution here. We fix the outer radius at $R_\mathrm{out} = 50$ $R_*$, which is sufficiently large such that the intensity has fallen off to less than 1\% of its starting value (i.e., a larger outer radius will have little effect on the final emission profile).

The stellar and disc emission profiles are added together and convolved with the COS LSF. In order to perform the fit, we conduct a Markov Chain Monte Carlo (MCMC) simulation using Python's \textsc{emcee} (\citealt{emcee}), to recover the overall radial velocity (RV -- the stellar and disc components are assumed to have the same RV), $v \sin i$, inner disc radius ($R_\mathrm{in}$) and inclination angle ($i_\mathrm{disc}$) of the disc (Table \ref{tab:mods}). We use 20 walkers for a total of 10,000 steps, the first 200 of which are discarded as burn-in. The starting position of each walker is sampled from a normal distribution centered on a chosen start value with a width of one-tenth of this value. We visually confirm that the walkers are exploring the parameter space sufficiently. We impose flat and relatively wide priors only (Table \ref{tab:mods}), as we do not have a good sense for constraints on these parameters for this system. Our final fitted parameters are also given in Table \ref{tab:mods}.

\begin{table}[hbt!]
    \caption{MCMC disc fit parameters}
    \begin{center}
    \begin{tabular}{l c c}
        \hline \hline
         Parameter & Flat Prior & Fit Value  \\
         \hline
         $v \sin i$ (km s$^{-1}$) & 0 -- 30 & 18.67$^{+3.10}_{-3.15}$ \\
         RV (km s$^{-1}$) & 0 -- 30 & 12.30 $^{+0.99}_{-1.06}$ \\
         $i_\mathrm{disc}$ ($^\circ$) & 0 -- 90 & 70.81$^{+13.14} _{-16.03}$ \\
         $R_\mathrm{in}$ ($R_*$) & 1 -- 10 & 3.58$^{+1.89}_{-1.32}$ \\
         \hline
    \end{tabular}
    \end{center}
    \tablecomments{To calculate emission radii (Table \ref{tab:comps}), we instead use an inclination angle of 80$^\circ$ (see text for details).}
    \label{tab:mods}
\end{table}


\subsection{The Optical Spectrum}
\label{subsec:optical}

\begin{figure*}[hbt!]
    \centering
    \includegraphics[scale=0.5]{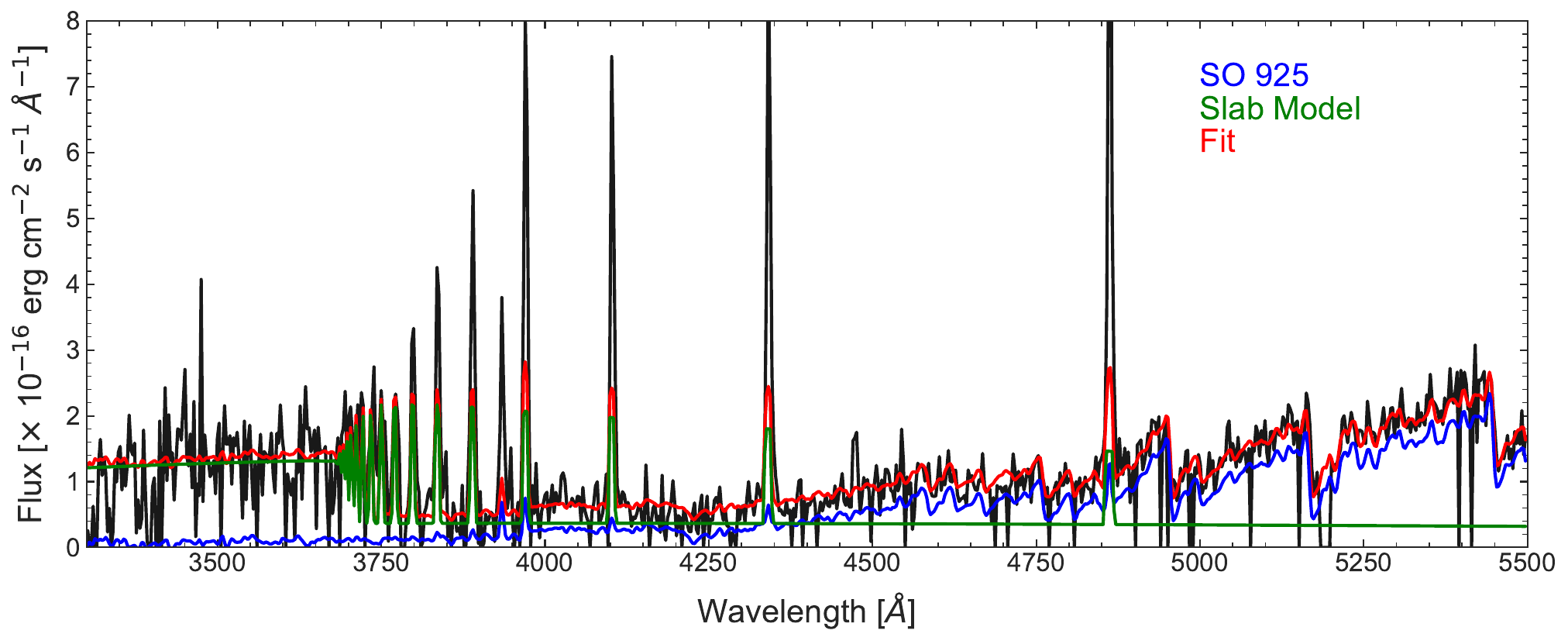}
    \caption{The blue curve is a non-accreting PMS template, SO 925. The green curve is the emission from the hydrogen slab model (see text for details), and the red curve is the sum of these two curves and the fit to the data. While the emission lines produced by the model appear to be a poor fit to the observation, this is simply because the slab model assumes that any hydrogen emission comes only from the slab. In reality, there is also hydrogen emission coming from the star itself, which is why the emission lines in the observation are stronger.}
    \label{fig:accmodel}
\end{figure*}

The mass accretion rate of young objects plays an important role in their evolution and can be estimated using the optical spectrum. We utilise optical data for J0844 from STIS that is nearly simultaneous with the COS FUV data. The accretion rate is estimated by modelling the observed spectrum with a stellar component added to an accretion component in a way analogous to \cite{rugel_x-shooter_2018}. Following \cite{rugel_x-shooter_2018}, we use a template spectrum, scaled by a multiplicative free parameter to account for differences in distance and radius, of a PMS non-accreting source of similar spectral type \citep[SO 925 -- spectrum taken from][]{manara2013} as a measure of the photospheric emission. We then add to it the spectrum of a slab of hyrodgen gas, assumed to be in local thermodynamic equilibrium (LTE), to quantify the accretion-heated regions of the star using the code of \citet[][see also \citealt{manara2021}]{valenti1993}. We fit the combined model to the observed optical spectrum of J0844. The fit parameters are calculated by minimising a $\chi^2$ function over the optical spectrum. From the resulting parameters, we compute the slab spectrum over an extended wavelength range and integrate to get the accretion luminosity, which is used to find the accretion rate $\dot{M}_\mathrm{acc}$ using
\begin{equation}
    \dot{M}_\mathrm{acc} = \frac{L_\mathrm{acc} R_*}{G M_*}\Biggl(1 - \frac{R_*}{R_{\text{in}}} \Biggr)^{-1} 
    \label{eq:lacc}
\end{equation}
It is often assumed that the disc inner (truncation) radius, $R_\mathrm{in}$, is approximately 5 $R_*$, close to the expected co-rotation radius (\citealt{hartmann_accretion_2016}). Using our measured rotation period (Section \ref{subsec:params}), we estimate the co-rotation radius for J0844 to be 4.86 $R_*$. Therefore, we use $R_\mathrm{in} = 5$ $R_*$ for the calculation of the mass accretion rate, though we explore $R_\mathrm{in}$ more in Section \ref{subsec:warmgas}. The fit parameters are shown in Table \ref{tab:modelfitparams}. For J0844, we estimate an accretion rate of $4.2 \times 10^{-11}$ $M_\odot$ $\mathrm{yr}^{-1}$. 

\begin{table}[hbt!]
    \centering
    \caption{Fit parameters of the hydrogen slab model shown in Figure \ref{fig:accmodel}.}
    \begin{tabular}{l  c}
    \hline \hline
        Parameter & Fit Value \\
        \hline
         Temperature (K)& 10,480 \\
         Optical depth at 3,500.4 \AA\ & 0.758 \\
         Optical depth at 4,000.2 \AA\ & 0.197 \\
         $L_\mathrm{acc}$ ($\times 10^{29}$ erg s$^{-1}$) & 4.87 \\
         $\dot{M}_\mathrm{acc}$ ($\times 10^{-11}$ $M_\odot$ yr$^{-1}$) & 4.2 \\
         \hline
    \end{tabular}
    \label{tab:modelfitparams}
\end{table}

\begin{figure}
    \centering
    \includegraphics[width=\linewidth]{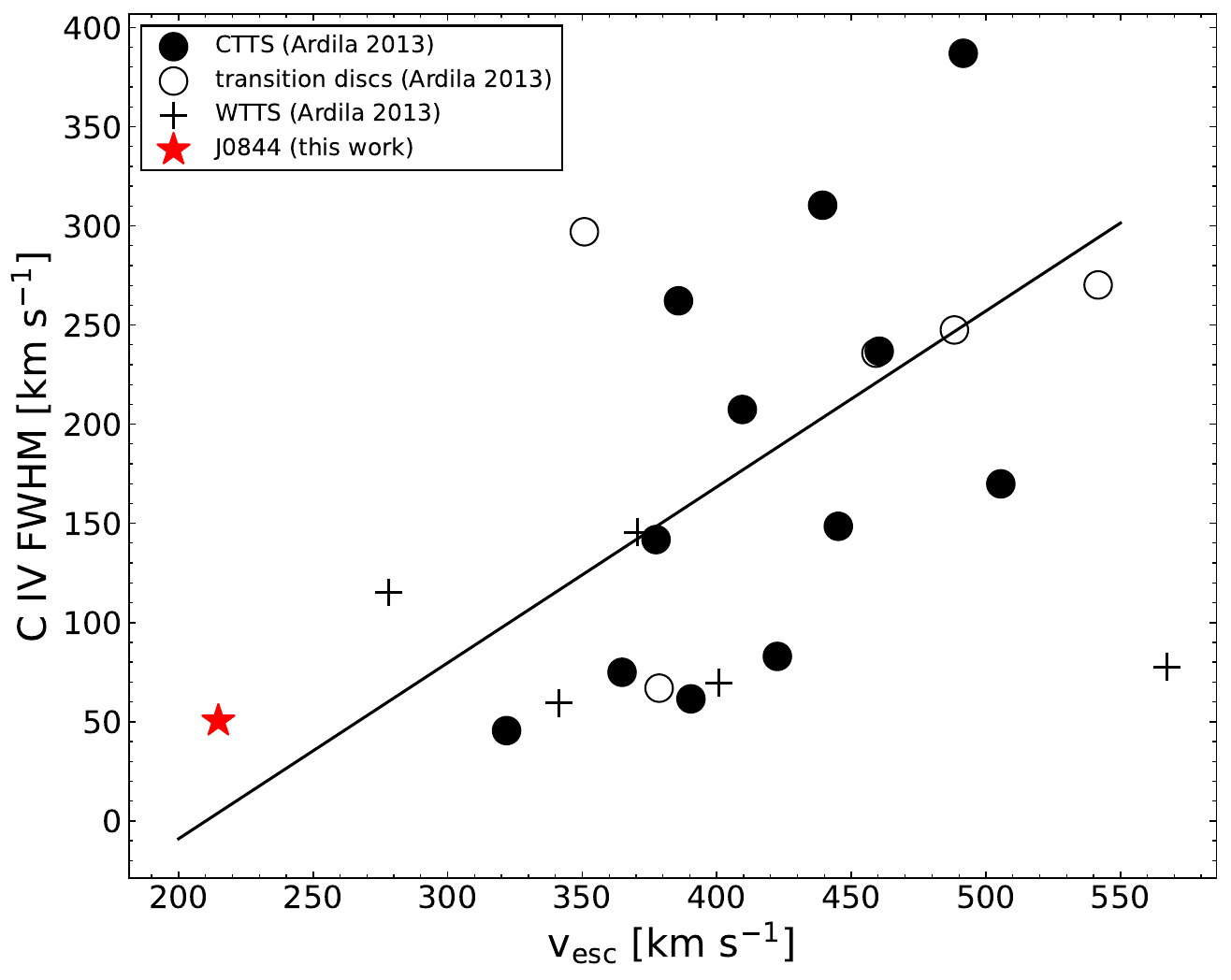}
    \caption{FWHM of the C \textsc{\footnotesize{IV}} lines versus escape velocity. The stellar masses are taken from \cite{france_hubble_2012}, and all FWHM values (except J0844) are from the \cite{ardila_hot_2013} non-parametric fits in their Table 5. J0844 is shown as a red star. Its FWHM values are also calculated non-parametrically (see \citealt{ardila_hot_2013} and the text) for both lines of the C \textsc{\footnotesize{IV}} doublet in Table \ref{tab:linefluxes}. The final reported and plotted value is the average of these. The line represents a least-squares regression fit to only the accreting objects (i.e., CTTSs and transition discs) excluding J0844. We find a Pearson correlation coefficient of $R = 0.53$ with $p = 0.03$.}
    \label{fig:placeholder}
\end{figure}

\begin{figure*}[hbt!]
    \begin{center}
    \includegraphics[scale=0.6]{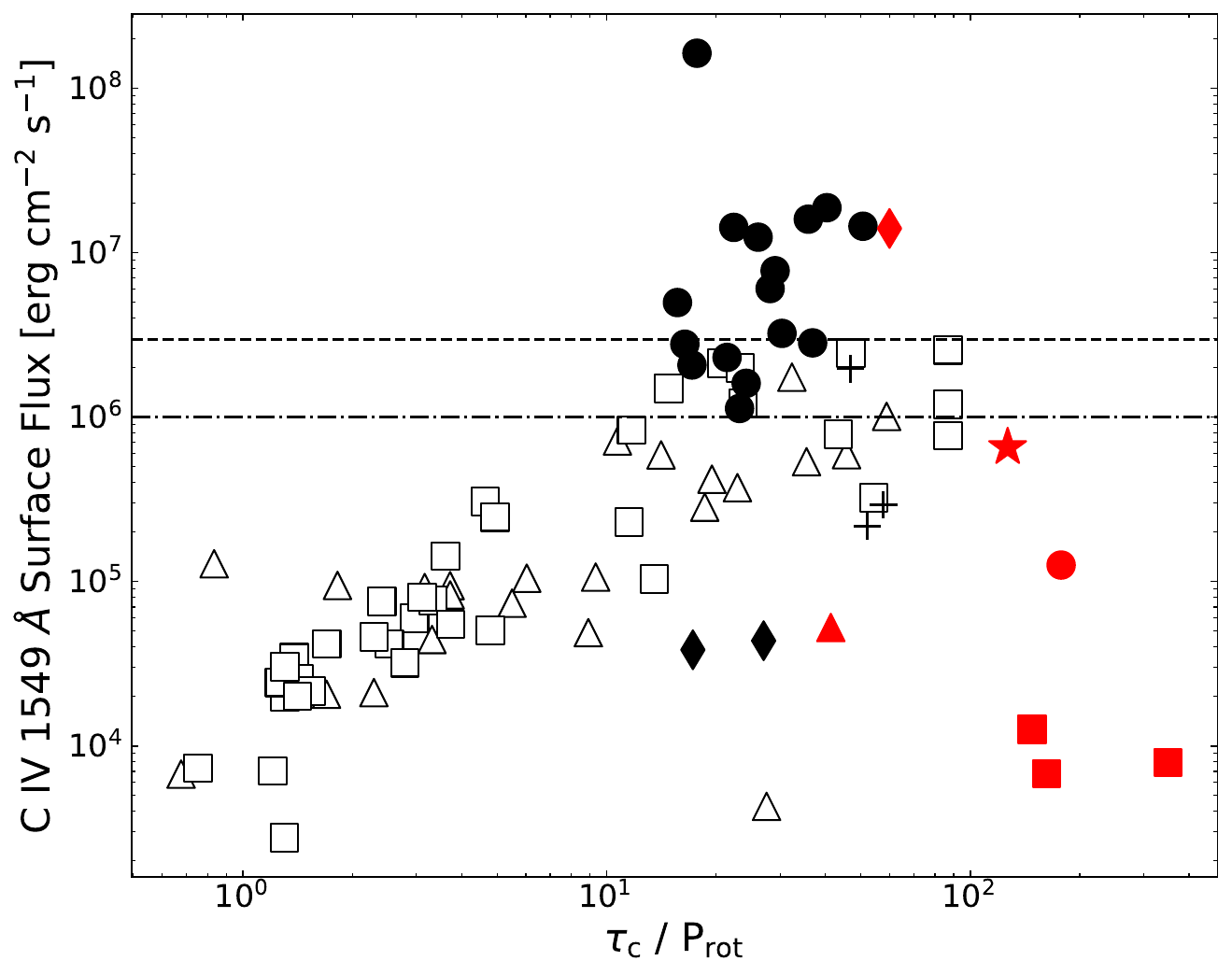}  
    \end{center}
    
    \caption{C \textsc{\footnotesize{IV}} surface flux vs. dynamo number ($\tau_c$/$P_\mathrm{rot}$). Adapted from \cite{jk_2000}. The solid circles are classical T Tauri stars (CTTSs), the crosses are weak-lined T Tauri stars (WTTSs), the squares are solar type main-sequence dwarfs, and the triangles are RS CVn stars. The two fully convective M dwarfs AD Leo and EV Lac are shown as solid diamonds. The active late M type very low mass stars VB 8, VB 10, and LHS 2065 (\citealt{hawley_transition_2003}) are shown as red squares. The red triangle is also EV Lac, but using the measured flux from \cite{osten_radio_2006} instead of \cite{jk_2000}, and using the period found in Section \ref{subsec:params}. The accreting brown dwarfs J1207 (\citealt{france_metal_2010}) and J0414 (\citealt{yang_far-ultraviolet_2012}) are shown as a red circle and red diamond respectively. J0844 (this work) is the red star. The dashed line is the dwarf star saturation level, and the dashed-dotted line is the TTS saturation level.} 
    \label{fig:civ}
\end{figure*}

\section{Results and Discussion} \label{sec:discussion}



\subsection{Hot Gas Tracers} \label{subsec:hotgas}

In accreting stars, hot FUV lines are expected to form close to the accretion shock where the material is heated sufficiently (e.g., $10^5$ K for C \textsc{\footnotesize{IV}}) to create high ionisation states. However, C \textsc{\footnotesize{IV}} is also produced through magnetic activity, so the main challenge in using these FUV lines as accretion tracers is separating emission originating from the transition region and emission originating in the accretion flows. One feature of accreting objects is a broad component of the C \textsc{\footnotesize{IV}} line profile, which we do recover for J0844 (see Section \ref{subsec:fuv}). 

However, for J0844, the C \textsc{\footnotesize{IV}} flux profiles (Figure \ref{fig:civfit}) are narrow compared to those generally associated with CTTSs, and are more akin to WTTSs. We find a BC FWHM of the two C \textsc{\footnotesize{IV}} lines equal to 111 $\pm$ 41 km s$^{-1}$ by taking the weighted mean of both lines widths. This is similar (within uncertainties) to the widths of several WTTSs in \cite{ardila_hot_2013}, which lie between about 80 and 100 km s$^{-1}$. In contrast, the CTTSs in \cite{ardila_hot_2013} generally have BC FWHMs of between 100 and 400 km s$^{-1}$ with only a few exceptions.

The relative narrowness of J0844's C {\footnotesize{IV}} lines is illustrated in Figure \ref{fig:placeholder} where we plot the C \textsc{\footnotesize{IV}} FWHM obtained through non-parametric fitting by \citet[][Table 5]{ardila_hot_2013} against the escape velocity of each source.
We follow the same method as \cite{ardila_hot_2013} to calculate the C \textsc{\footnotesize{IV}} FWHM of J0844 for both lines of the doublet. Because of the relatively low signal-to-noise in the J0844 spectra, we use the peak values of the fits (red lines in Figure \ref{fig:civfit}) as the line maxima and measure the FWMM based on this value, interpolating between wavelength pixels as needed. We estimate an uncertainty by finding the FWHM between the pixels on either side of our interpolation. We then average the widths of the two lines. We find an average FWHM of 50.41 $^{+3.2}_{-7.7}$  km s$^{-1}$.  

We plot these FWHM values against the escape velocity which is equal
to the free-fall velocity if one assumes material falling in from infinity. If material falls
in from the disc truncation radius ($R_\mathrm{in}$), the velocity reached by the accreting material is decreased. However, if we assume $R_\mathrm{in}/R_*$
is the same for all objects (usually assumed equal
to 5; \citealt{hartmann_accretion_2016}) then the velocity reached by the accreting material is a constant fraction (0.89) of the escape velocity for all objects.  We perform a least-squares regression fit for the CTTSs and transition disc objects (ignoring J0844) and find a Pearson correlation coefficient of 0.53, statistically significant at the 97\% level, indicating a relatively strong correlation between the C \textsc{\footnotesize{IV}} line width and the escape velocity.

The correlation seen in Figure \ref{fig:placeholder} suggests the width of the C {\footnotesize{IV}} lines is due in part to Doppler broadening produced by the magnetospheric flow in accreting objects.  On the other hand, the WTTSs show consistently low FWHM values and reveal no correlation with escape velocity as expected for objects not undergoing accretion.  The narrowness of J0844's C {\footnotesize{IV}} lines do not clearly indicate these lines form in an accretion flow.

One possibility is that the C \textsc{\footnotesize{IV}} in J0844 is caused entirely by magnetic activity, which is the case for WTTSs. However, the presence of its circumstellar disc (\citealt{megeath_spitzerirac_nodate}) and excess blue continuum emission (\citealt{rugel_x-shooter_2018} and Figure \ref{fig:accmodel} of this work) indicates that J0844 is accreting and was indeed accreting at the time that the C \textsc{\footnotesize{IV}} observations were made. Therefore, we expect that at least some of the C \textsc{\footnotesize{IV}} flux is produced by the accretion shock.

The challenge is separating how much each process is contributing to the observed emission. \cite{jk_2000} established that CTTSs show levels of C \textsc{\footnotesize{IV}} at or above the activity saturation level for non-accreting objects (Figure \ref{fig:civ}), and showed that this excess flux is well-correlated with the accretion rate. If we place J0844 and J1207 on the same plot, using $\tau_\mathrm{c}$ and $P_\mathrm{rot}$ from Section \ref{subsec:params}, both indeed lie in the saturated regime. However, both targets show levels of C \textsc{\footnotesize{IV}} emission well below what is expected from magnetic activity alone, even though both objects are thought to be accretors. We argue that in the brown dwarf regime, we need to redefine the baseline for magnetic activity saturation, i.e., the point from which the ``excess'' is calculated. 

To get a sense of the magnetic activity level that can be expected at these low masses, we consider the C \textsc{\footnotesize{IV}} emission from the late active M-type stars VB 8, VB 10, and LHS 2065 (\citealt{hawley_transition_2003}). We again assume a convective turnover time of 180 days, and use the periods calculated in Section \ref{subsec:params}. This places all three objects in the saturated regime. We also include the earlier active M-type star, EV Lac (\citealt{osten_radio_2006}), using $\tau_\mathrm{c}$ = 180 days and the period found in Section \ref{subsec:params} for this object. 

If we compare J0844 and J1207 to these objects, both show excess C \textsc{\footnotesize{IV}} emission, in the same way that higher mass CTTSs show excess C \textsc{\footnotesize{IV}} emission compared to their non-accreting counterparts. In Section \ref{subsec:acc} below, we calculate an accretion rate based on this excess C \textsc{\footnotesize{IV}} luminosity.

\begin{figure}[hbt!]
    \centering
    \includegraphics[scale=0.35]{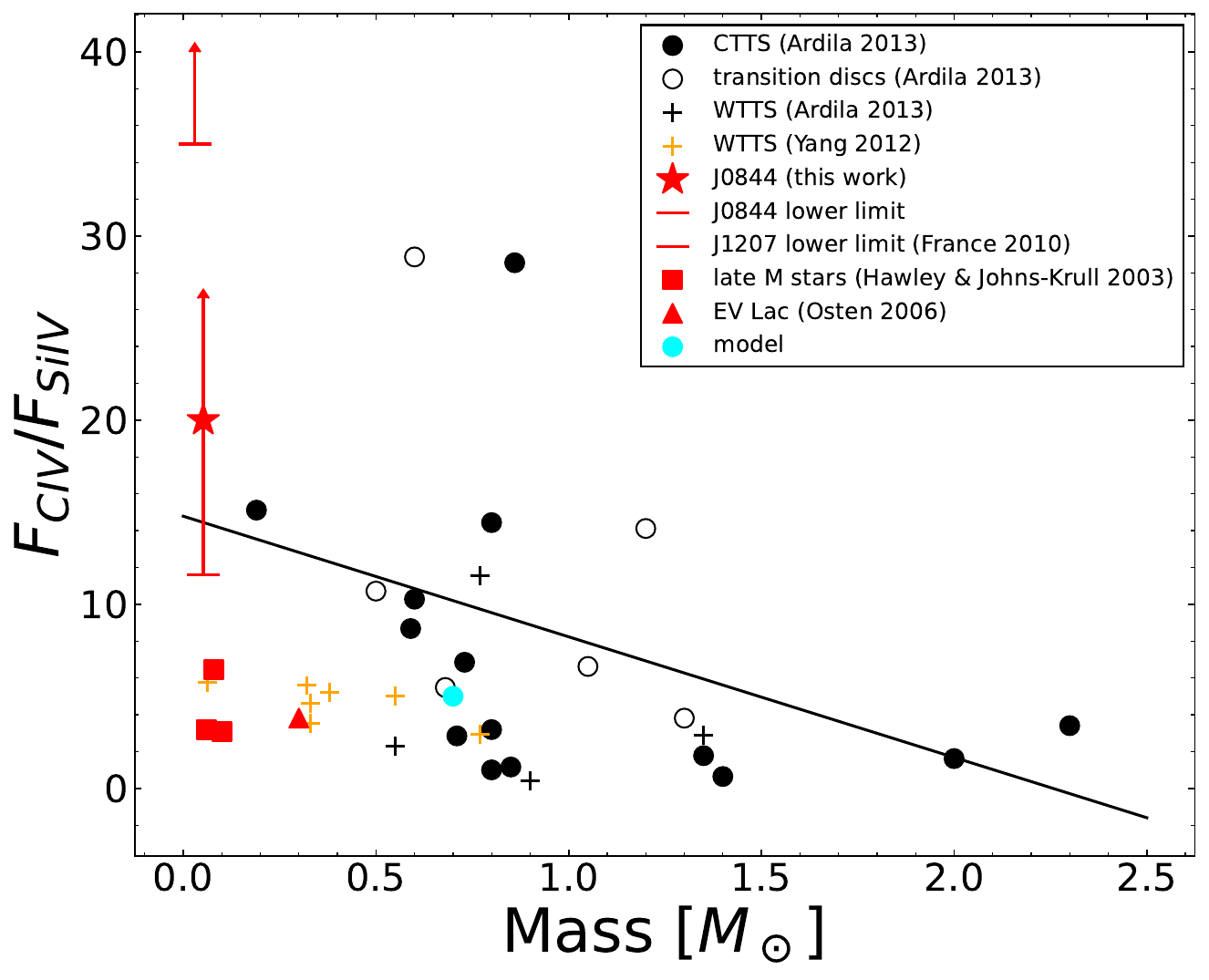}
    \caption{Ratio of C \textsc{\footnotesize{IV}} to Si \textsc{\footnotesize{IV}} flux, as a function of stellar mass. The stellar masses are taken from \cite{france_hubble_2012}, while the flux measurements are from \cite{ardila_hot_2013}. J0844, the red star, is shown as a lower limit (see text for details). The line represents a least-squares regression fit to only the accreting objects, i.e., the CTTSs, transition discs, J0844, and J1207. The cyan filled circle represents a simple shock model applied to a ``baseline'' CTTS, as described in the text. We find a Pearson correlation coefficient of $R=-0.4$ with $p=0.08$.}
    \label{fig:ratio}
\end{figure}

\begin{figure}
    \centering
    \includegraphics[width=\linewidth]{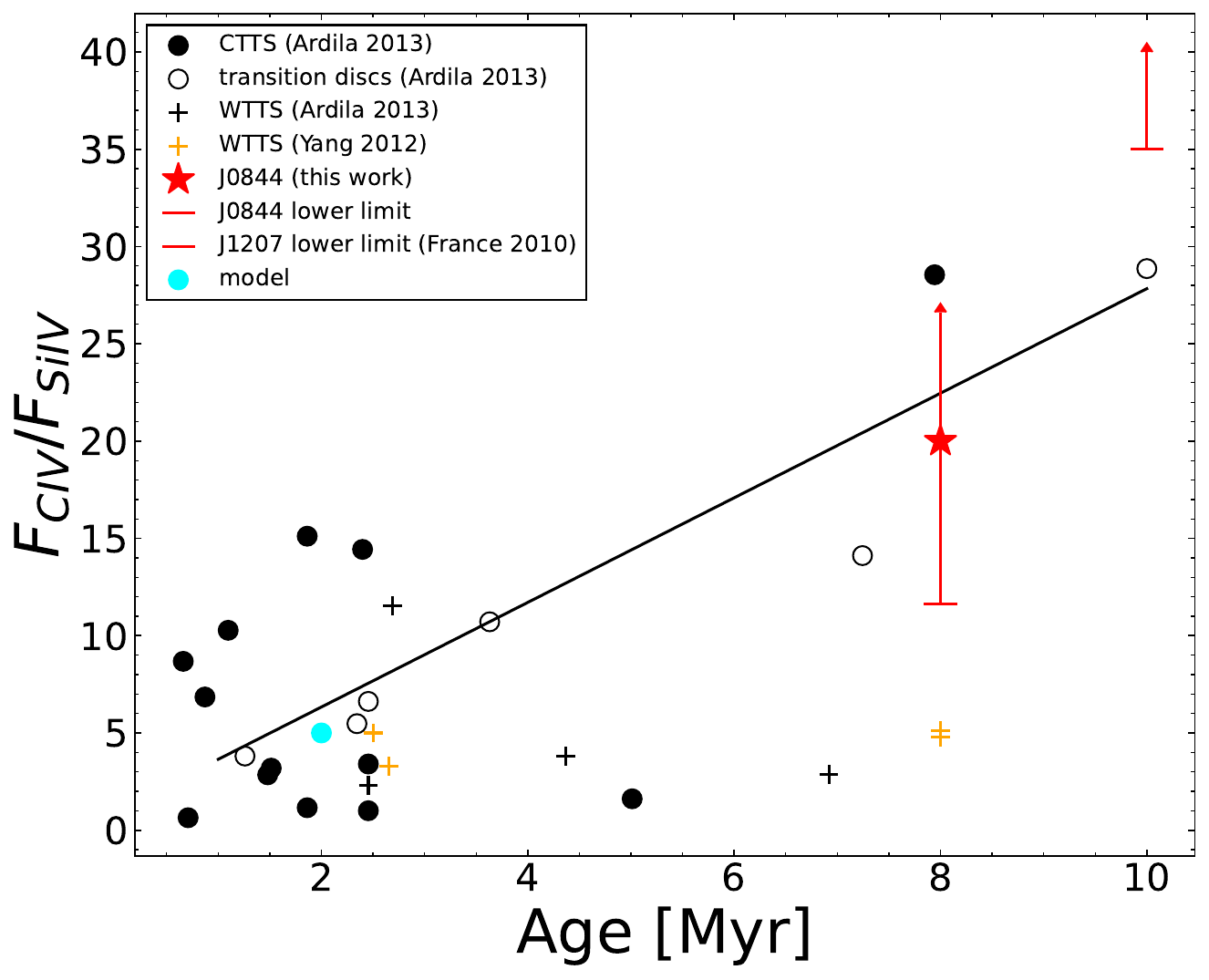}
    \caption{Ratio of C \textsc{\footnotesize{IV}} to Si \textsc{\footnotesize{IV}} flux, as a function of age. The ages are again taken from \cite{france_hubble_2012}, with the flux measurements from \cite{ardila_hot_2013}. We choose an age of 2 Myr for our ``baseline'' CTTS (cyan filled circle). As in Figure \ref{fig:ratio}, we only perform the linear least squares regression on the accreting objects. We find a Pearson correlation coefficient of $R=0.84$ with $p=2\times 10^{-6}$.}
    \label{fig:ratio_age}
\end{figure}

While much of our focus has been on C \textsc{\footnotesize{IV}}, we also examine other hot gas lines. In particular, we examine the Si \textsc{\footnotesize{IV}} to C \textsc{\footnotesize{IV}} ratio, as this has been used as an indicator of accretion \citep[e.g.,][]{herczeg_farultraviolet_2002}. Since C \textsc{\footnotesize{IV}} and Si \textsc{\footnotesize{IV}} form at similar temperatures, the C \textsc{\footnotesize{IV}}/Si \textsc{\footnotesize{IV}} flux ratio is a proxy for the overall C/Si ratio, which is what we refer to in the following. The relative strengths of the C and Si emission give information about the composition of the gas in the accretion column and the extent to which dust-processing has occurred: a high C/Si ratio indicates significant dust-processing, locking the Si up in grains and depleting the accretion flow of Si gas \citep{furlan_spitzer_2005}. On the other hand, a lower C/Si ratio indicates that most the Si is in the gas phase, and therefore free to emit.
For J0844, the C/Si ratio is found to be 20 when adding the fluxes from the two doublet components in Table \ref{tab:params} together. We also calculate a lower limit on this ratio by assuming all of the emission near $\lambda$1393,1402 \AA\ is Si \textsc{\footnotesize{IV}} (i.e., no H$_2$ contribution). We find this lower limit ratio to be 11. \cite{pittman2025} find a C \textsc{\footnotesize{IV}}/Si \textsc{\footnotesize{IV}} ratio of 10.7 using fluxes measured by \cite{france_radial_2023} (who assume no H$_2$ contribution). This small difference stems from the fact that \cite{pittman2025} used an extinction value of A$_\mathrm{V} = 0.1$ mag found from their accretion shock modelling, whereas we use A$_\mathrm{V} = 0$ mag from \cite{rugel_x-shooter_2018}, highlighting the effects of uncertainties in extinction magnitudes on the value of this ratio. Figure \ref{fig:ratio} shows C/Si (C \textsc{\footnotesize{IV}}/Si \textsc{\footnotesize{IV}}) for a sample of CTTSs, transition disc objects, and WTTSs, taken from \cite{ardila_hot_2013} and \cite{yang_far-ultraviolet_2012}. We include only the WTTSs from \cite{yang_far-ultraviolet_2012}, because these were low resolution observations. CTTSs generally show H$_2$ right around the Si \textsc{\footnotesize{IV}} doublet, which we would be unable to reliably separate at low spectral resolution. This is why the accreting brown dwarf J0414 appears in Figure \ref{fig:civ}, but not in Figure \ref{fig:ratio} and \ref{fig:ratio_age}. 

Some accreting objects, such as TW Hya and V4046 Sgr, have C/Si ratios of around 30. If we calculate what the ratio should be for solar metallicity with all silicon in the gas phase using a simple shock model (see Section \ref{sec:lcivlacc} for details), we find a ratio of $\sim 5$ (cyan filled circle in Figure \ref{fig:ratio}), which agrees relatively well with the ratio for some accreting stars as well as for the WTTSs, the very low mass active stars (red
squares) from \citet{hawley_transition_2003}, and the active low-mass star 
EV Lac \citep[red triangle;][]{osten_radio_2006} shown in Figure \ref{fig:ratio}. 
J0844's ratio is higher than EV Lac and some TTSs, as well as other active M stars, but still lower than objects such as J1207, TW Hya, and V4046 Sgr. This suggests that at least some of J0844's Si has been processed into silicates and lost to the accreting gas (\citealt{furlan_spitzer_2005}), though not as much as for the targets with higher ratios. The silicates will probably have settled to the midplane of the disc, which is consistent with \cite{apai_onset_2005}, who find evidence for dust processing and grain growth in several young brown dwarf candidates.

Since the C/Si ratio is often used as an accretion diagnostic (as mentioned above), it is also interesting to note that there does not appear to be a significant difference between the ratios for many of the CTTSs relative to the WTTSs. This suggests that perhaps the C/Si ratio is not an entirely reliable indicator of accretion in and of itself, though more in-depth modelling of the production of these lines is needed.

It is interesting to note that two of the objects with the highest C/Si ratio are two of the lowest mass objects. Indeed, the
modelling of \citet{mah2023} predicts that lower mass stars should have a
higher abundance of C relative to refractory elements such as Si due
to the difference in timescales for accreting gas (the viscous timescale)
relative to accreting pebbles which accrete much faster due to drift 
caused by gas drag. Statistically though, the correlation of the C/Si
ratio with mass is weak, with a correlation coefficient of $R = -0.4$ and
an associated false alarm probability of $p = 0.08$ using the Pearson correlation test. When calculating this correlation coefficient, we restrict 
ourselves to objects whose emission is thought to be dominated by accretion:
CTTSs, transition discs, and J0844 and J1207.
The depletion of elements such as Si appears to also correlate with age (Figure \ref{fig:ratio_age}). \cite{huhn2023} model stellar elemental abundances in star-disc systems over time. While they focus on Fe as a proxy for all refractory elements, the behaviour of Si relative to Fe should be similar given their similar condensation temperatures. We thus assume the C/Fe behaviour found by \cite{huhn2023} also applies to the C/Si ratio. They find that the C/Fe ratio in the accreting gas should increase with age due to the depletion of elements such as Fe (and Si) due to the rapid accretion of pebbles. Figure \ref{fig:ratio_age} is the same as Figure \ref{fig:ratio} but with the C/Si ratio plotted against age instead. The line indicates a least-squares regression fit. To perform the fit, we again use only the likely accreting objects: the CTTSs, transition disc objects, J0844, and J1207. We see that the C/Si ratio increases with age, with a relatively strong Pearson correlation coefficient of $R = 0.84$ and small $p$-value ($2 \times 10^{-6}$). Thus, the C/Fe behaviour with age found by \cite{huhn2023} appears to agree well with our observed behaviour of C/Si with age, and we suggest it is primarily the age effect instead of the mass effect that leads to the oberved variation in the C/Si ratio observed in these accreting young objects.


\subsection{Warm Gas Tracer: \normalfont{H}$_2$} \label{subsec:warmgas}

The width of the broad component of the Gaussian fit to the H$_2$ lines can be used to estimate both the inner disk truncation radius as well as the average radius at which the H$_2$ emission originates. We calculate the inner H$_2$ radius $R_\mathrm{in}(\mathrm{H}_2)$, and the average H$_2$ radius $\langle R_{\mathrm{H}_2} \rangle$ following \citet[][see Equation \ref{inH2} and \ref{avH2} below]{france_hubble_2012}.
We use a disc inclination angle of 80$^\circ$ based on the more physical disc fit as discussed below. We find $R_\mathrm{in}(\mathrm{H}_2) = 1.92$ $\pm$ $0.35$ $R_*$, 
and $\langle R_{\mathrm{H}_2} \rangle = 5.54$ $\pm$ $1.03$ $R_*$, where the uncertainties take into account the uncertainties in the HWHM. 

\begin{equation}
    R_\mathrm{in}(\mathrm{H}_2) = G M_* \Biggl(\frac{\sin i}{1.7 \times \mathrm{HWHM}_\mathrm{BC}} \Biggr)^2
    \label{inH2}
\end{equation}

\begin{equation}
    \langle R_{\mathrm{H}_2} \rangle = G M_* \Biggl(\frac{\sin i}{\mathrm{HWHM}_\mathrm{BC}} \Biggr)^2
    \label{avH2}
\end{equation}

When considering the more physical disc fit, the $v$ in a star's $v \sin i$ (Table \ref{tab:mods}) is the equatorial velocity, $v_\mathrm{eq}$, which can be calculated using the radius and rotation period of the star. Using the period we calculate in Section \ref{subsec:params} of $P = 1.42$ days, and a radius of $R_* = 0.43$ $R_\odot$, we recover $v_\mathrm{eq} = 15.32$ km s$^{-1}$. Of course, the $v \sin i$ must be lower than or equal to the equatorial velocity, and we find a $v \sin i$ of $18.67$ km s$^{-1}$ from the MCMC fit (Table \ref{tab:mods}). However, these two values agree within roughly 1$\sigma$. The high $v \sin i$ value favours a high inclination angle for the star. This is further supported by \cite{pittman2025} who find that J0844's magnetosphere is probably viewed edge-on ($i_\mathrm{mag} = 90^\circ$). Given that magnetic obliquities are typically 20$^\circ$ or less, this favours a high stellar inclination angle. The disc inclination angle need not be the same as the star. We find a disc inclination of $i_\mathrm{disc} = 70.81^\circ$. However, the spread in $i_\mathrm{disc}$ is large, and there is a correlation between $R_\mathrm{in}$ and $i_\mathrm{disc}$. This is unsurprising because these parameters are somewhat degenerate: increasing the inclination angle can be balanced by decreasing the inner radius, and vice versa. Given that it is unlikely that the inclination angle of the disc will differ dramatically from that of the star at these small radii ($<5$ $R_*$), it is likely that the disc has a high inclination angle as well. Therefore, we fix the disc inclination angle at 80$^\circ$, roughly 1$\sigma$ greater than the original fit value. Holding everything else constant as the best fit values in Table \ref{tab:mods}, we then fit only for the inner radius. This results in an inner radius of $R_\mathrm{in}(\mathrm{H}_2) = 3.28^{+2.32}_{-1.57}$ $R_*$, which is also within 1$\sigma$ of the original fit value. This value is similar to but larger than the value calculated in Section \ref{subsubsec:doublegauss} using the HWHM of the broad Gaussian component. We calculate an average emission radius using
\begin{equation}
    \langle R_{\mathrm{H}_2} \rangle = \frac{\int_1^{R_\mathrm{out}} r I(r) \mathrm{d} r}{\int_1^{R_\mathrm{out}} I(r) \mathrm{d} r}
    \label{eq:disch2ave}
\end{equation}
where $I(r)$ is given by Equation \ref{ieq}. This results in $\langle R_{\mathrm{H}_2} \rangle = 8.74$ $R_*$. In Table \ref{tab:comps}, we summarise the radii calculated using the two different methods.

\begin{table}[hbt!]
    \caption{\textbf{H$_2$ emission radii}}
    \begin{center}
        \begin{tabular}{l c c}
        \hline \hline
         Quantity &  Double Gaussian Fit & Star + Disc Fit \\
         \hline
         $R_\mathrm{in}(\mathrm{H}_2)$ ($R_*$) & $^*$1.92 $\pm$ 0.35  & $^\square$3.28$^{+2.32}_{-1.57}$  \\
         $\langle R_{\mathrm{H}_2} \rangle$ ($R_*$) & $^\dagger$5.54 $\pm$ 1.03 & $^\triangle$8.74 \\
         \hline
    \end{tabular}
    \end{center}
    
    \tablecomments{Inner and average H$_2$ radii were all calculated using a disc inclination of $80^\circ$, and the MCMC simulation was run with the disc inclination fixed at $80^\circ$ (see text for details). \\ $^*$Calculated according to Equation \ref{inH2}, assuming uncertainty only in HWHM. \\ $^\dagger$Calculated according to Equation \ref{avH2}, assuming uncertainty only in HWHM. \\ $^\triangle$Calculated according to Equation \ref{eq:disch2ave}. \\ $^\square$Fit parameter from the MCMC simulation (with $i_\mathrm{disc}$ fixed at 80$^\circ$).}
    \label{tab:comps}
\end{table}

For both fitting methods of the coadded H$_2$ line profile, we find an inner emission radius that lies inside the co-rotation radius [$R_\mathrm{in}(\mathrm{H}_2) = 1.92, 3.28$ $R_*$ for the double Gaussian and star-disc fits respectively, while $R_\mathrm{co} = $ 4.86 $R_*$]. Steady accretion generally occurs when the disc is truncated by the central object's magnetic field somewhere inside of the co-rotation radius \citep{hartmann_accretion_2016}, though highly episodic accretion can occur under certain circumstances in the propeller regime when the central object's angular velocity exceeds that of the inner disc \citep[e.g.,][]{Lii2014}. Exactly how close to the central object the disc is expected to be truncated depends on several factors, including the disc accretion rate as well as the strength of the central object's magnetic field. Simulations by \cite{romanova2008} identify two regimes of accretion. In the stable regime, the truncation radius is relatively close to the co-rotation radius, while in the unstable regime the truncation radius is significantly smaller than the co-rotation radius. For J0844, we believe the inner disc radius determined from the actual disc fit (Section \ref{subsubsec:disc}) to be the more accurate estimate, and this gives a truncation radius that is $\sim 0.7$ times the co-rotation radius. In their study of inner discs around accreting young stars as traced by CO rovibrational emission, \cite{salyk_co_2011} find that for CTTSs, the inner disc typically survives to radii of $0.5 - 1.0$ times the truncation radius, nicely bracketing the value we get for J0844.

It is worth considering how much of the H$_2$ emission may originate in the accretion flow itself, which we do not account for in our model. Knowledge of the temperature structure of the accretion flow is generally quite limited, and how this depends on the central object's mass is not well known. Models of magnetospheric accretion that try to reproduce Balmer emission lines often find peak temperatures in the accretion flow of $\sim 8000$ K (e.g., \citealt{muzerolle2001}), while similar models applied to accreting very low mass stars and brown dwarfs have peak temperatures of 12,000 K (\citealt{muzerolle2005}); however, it remains a mystery just how the material is heated to these temperatures. For example, the self-consistent thermal models of CTTS magnetospheres of \cite{martin1996} only reach a maximum temperature of $\sim 6000$ K, so the maximum temperature in the magnetosphere is usually treated as a free parameter (e.g., \citealt{muzerolle2001, wilson2022}). If the outer parts of the magnetosphere of J0844 remain relatively cool, it is possible that some of the H$_2$ emission we observe could form in this region. However, if the magnetosphere is rotating with the star, the velocity in this region will be sub-Keplerian and will not contribute to the largest observed projected velocities. One thing that seems certain from our analysis of the H$_2$ emission line is that the disc does not extend down to the surface of J0844. The stellar temperature is low enough for H$_2$ to be relatively abundant, so the disc just above the stellar surface and outside any potential boundary layer should also be cool enough to contribute H$_2$ emission. The fact that we observe an inner radius for the disc contribution to the H$_2$ emission likely indicates the disc is indeed truncated, presumably by the magnetic field of J0844, well before it reaches the stellar surface.

\subsection{Accretion}
\label{subsec:acc}

In Table \ref{tab:maccs}, we list the various mass accretion rate estimates for J0844 from the literature and from this work, including how each value was calculated. 

Our mass accretion rate for J0844 from the optical continuum ($4.2\times 10^{-11}$ $M_\odot$ yr$^{-1}$) is about two thirds that of \cite{rugel_x-shooter_2018}'s optical continuum estimate, probably due to the fact that the overall blue part of the spectrum is about two thirds as strong in our data compared to \cite{rugel_x-shooter_2018}. This is not too surprising given the typical variability of young accreting stars \citep[e.g.,][]{herczeg_twenty-five_2023}. Our mass accretion rate from the optical continuum is also fairly consistent with \cite{rugel_x-shooter_2018}'s estimates using H$\alpha$ and H$_\beta$ and the stellar scaling relationships from \cite{alcala_x-shooter_2014}. We note, though, that \cite{rugel_x-shooter_2018}'s mass accretion rate based on H$\alpha$ is about two and a half times lower than their mass accretion rate calculated from the optical continuum, whereas our mass accretion rate calculated from H$\alpha$ is nearly identical to that calculated from the optical continuum.

\cite{hashimoto2025} calculated mass accretion rates using H$\beta$ and either the stellar scaling relationships from \cite{alcala2017}, or the predictions from the shock model in \cite{aoyama_comparison_2021}. Our mass accretion rate calculated from the continuum excess and from H$\beta$ agree relatively well with \cite{hashimoto2025} when the stellar scalings from \cite{alcala2017} are used, but are about half (or less) of the accretion rates found by \cite{hashimoto2025} when using the shock model from \cite{aoyama_comparison_2021}. We note that \cite{rugel_x-shooter_2018} and \cite{hashimoto2025} analyse the same X-Shooter spectrum and get slightly different accretion rate estimates based on H$\beta$. This is because they used slightly different stellar parameters, and \cite{rugel_x-shooter_2018} used the scaling relations from \cite{alcala_x-shooter_2014}, while \cite{hashimoto2025} used updated scaling relations from \cite{alcala2017}. When we use excess C \textsc{\footnotesize{IV}} (and the relation from \cite{jk_2000}; see Equation \ref{eq:macc}) to calculate a mass accretion rate, we recover a value 1-2 orders of magnitude higher than any of the other mass accretion rate estimates in Table \ref{tab:maccs}, which we discuss further in Section \ref{sec:lcivlacc}. All of these findings together further reaffirm that the stellar scaling relationships (both in the optical and in the FUV) may be much more uncertain as one moves into the brown dwarf and planetary regime.


\cite{pittman2025} also analyse the same \textit{HST}-STIS spectra that we analyse here. While we analyse only the G430L spectrum of J0844, \cite{pittman2025} simultaneously fit the G230L, G430L, and G750L with the multi-component accretion shock models described in \citet{robinson2019} and \citet{pittman2022}. In the case of J0844 (identified as RECX16 in \citealt{pittman2025}), they find only two accretion column components are needed to fit the data, resulting in a mass accretion rate estimate of $5 \times 10^{-11}$ $M_\odot$ yr$^{-1}$, very close to our value of $4.2 \times 10^{-11}$ $M_\odot$ yr$^{-1}$. This difference is due to the stellar parameters that they adopt compared to ours, since quantities like the stellar mass and radius, and the disc truncation radius, enter the estimate of the accretion rate through Equation \ref{eq:lacc}. \cite{pittman2025} adopt/fit the following parameters for J0844: spectral type of M6, $T_\mathrm{eff} = 2770$ K, $M_* = 0.05$ $M_\odot$, $R_* = 0.49^{+0.08}_{-0.07}$ $R_\odot$, and an inner truncation radius of $3.83 \pm 0.25$ $R_*$. They use essentially the same mass that we do. If we use their values for the radius and inner truncation distance, our accretion rate estimate would become $5.2 \times 10^{-11}$ $M_\odot$ yr$^{-1}$, almost identical to that found by \cite{pittman2025}.
For the disc inner truncation radius, we simply use the commonly adopted value of 5 $R_*$, while \cite{pittman2025} fit the observed H$\alpha$ emission profile with magnetospheric accretion models to derive a value of 3.83$\pm 0.25$ $R_*$.  We note that the inner truncation radius we derive from our more physical stellar plus full disc model (rather than the double Gaussian fit) to the H$_2$ emission profile is 3.28 $R_*$, fully consistent with \cite{pittman2025} within uncertainties. It is likely then that the inner disc of J0844 is truncated somewhat interior to 5 $R_*$.

\subsection{$L_\mathrm{CIV}/L_\mathrm{acc}$}
\label{sec:lcivlacc}

Comparing J0844 to J1207, \cite{france_metal_2010} found an accretion rate of $6.6 \times 10^{-11}$ $M_\odot$ yr$^{-1}$ for J1207 based on its C \textsc{\footnotesize{IV}} luminosity, whereas we find an accretion rate of $4.2 \times 10^{-11}$ $M_\odot$ yr$^{-1}$ for J0844 based on the optical continuum. This is surprising, because J0844 shows a significantly larger amount of C \textsc{\footnotesize{IV}} emission compared to J1207 (Figure \ref{fig:civ}). \cite{france_metal_2010}'s accretion estimate was found by extrapolating the $L_\mathrm{CIV}$--$\dot{M}_\mathrm{acc}$ relationship for higher mass stars in \cite{jk_2000}, given by
\begin{equation}
    \log_{10}(\dot{M}_\mathrm{acc}) = 0.753 \log_{10}(L_{\mathrm{CIV,ex}}) - 29.89
    \label{eq:macc}
\end{equation} 
where $L_\mathrm{CIV,ex}$ is the excess luminosity observed compared to non-accreting but still magnetically active sources. \cite{france_metal_2010} assumed that all of J1207's C \textsc{\footnotesize{IV}} emission results from accretion processes, due to its weak magnetic field.

Using this same method, and taking VB 8, VB 10, and LHS 2065 as approximate measures of C \textsc{\footnotesize{IV}} coming from magnetic activity at these low masses, we adopt a surface flux saturation level of 10$^4$ erg cm$^{-2}$ s$^{-1}$ in Figure \ref{fig:civ}. This implies that J0844 has an excess C \textsc{\footnotesize{IV}} luminosity of $L_\mathrm{CIV,ex} = 5.7 \times 10^{27}$ erg s$^{-1}$. If we use this value in Equation \ref{eq:macc}, we get a mass accretion rate of $\dot{M}_\mathrm{acc} = 1.02 \times 10^{-9}$ $M_\odot$ yr$^{-1}$ for J0844, which is nearly two orders of magnitude higher than our continuum estimate. Even if we use Equation (1) from \cite{jk_2000}, which yields a lower accretion rate of $2.3 \times 10^{-10}$ $M_\odot$ yr$^{-1}$, there is still nearly an order of magnitude difference. A similar result was found by \cite{pittman2025} who similarly deduce that the C \textsc{\footnotesize{IV}} luminosity of J0844 is about two orders of magnitude stronger than expected based on a fit of the accretion luminosity as a function of C \textsc{\footnotesize{IV}} luminosity for their sample of accreting young objects.  This likely explains the discrepancy between the accretion rates and C \textsc{\footnotesize{IV}} fluxes of J1207 and J0844: the use of C \textsc{\footnotesize{IV}} with the relationships in \cite{jk_2000} may simply overestimate the mass accretion rate at these low masses.

\begin{table*}[hbt!]
    \caption{Comparison of accretion rate estimates for J0844}
    \begin{center}
    \begin{tabular}{c c c c c}
        \hline
        Observation Date & Instrument & Method & $\dot{M}_{\text{acc}}$  & Reference \\
        & & & ($\times 10^{-11}$ $M_\odot$ yr$^{-1}$) & \\
        \hline \hline
        2010-01-18 & VLT/X-Shooter & Continuum excess & 6.6 & \cite{rugel_x-shooter_2018} \\
        2010-01-18 & VLT/X-Shooter & H$\alpha$ & 2.7 & \cite{rugel_x-shooter_2018} \\
         2010-01-18 & VLT/X-Shooter & H$\beta$ & 5.5 & \cite{rugel_x-shooter_2018} \\
         2010-01-18 & VLT/X-Shooter & H$\beta$ & $^\triangle$5.6, 13.8 & \cite{hashimoto2025} \\
         2021-04-27 & VLT/X-Shooter & H$\beta$  & $^\triangle$2.4, 7.2 & \cite{hashimoto2025}\\
         2021-05-01 & VLT/X-Shooter & H$\beta$  & $^\triangle$3.6, 9.7 & \cite{hashimoto2025} \\
         2021-06-10 & HST/STIS & $^*$Continuum excess & 5.0 & \cite{pittman2025} \\
         2021-06-10 & \textbf{HST/STIS} & Continuum excess & 4.2 & This work \\
         2021-06-10 & HST/STIS & H$\alpha$ & 4.1 & This work \\
         2021-06-10 & HST/STIS & H$\beta$ & 5.4 & This work \\
         2021-06-10 & HST/COS & $^\dagger$Excess \textsc{C} \textsc{\footnotesize{IV}} & 102 & This work \\ 
         \hline
    \end{tabular}
    \end{center}
    \tablecomments{$^\triangle$ The first value is calculated by converting  $L_{\mathrm{H}\beta}$ to $L_\mathrm{acc}$ using the stellar scaling relationship from \cite{alcala2017}. The second value is calculated in the same way, except here the authors use predictions from the shock model from \cite{aoyama_comparison_2021}. For both accretion rate values, we use the $L_\mathrm{acc}$ values from \cite{hashimoto2025}, and then calculate $\dot{M}_\mathrm{acc}$ using Equation \ref{eq:lacc} and the parameters in Table \ref{tab:params}. $^*$The entire optical spectrum was modelled using a multi-component accretion shock model, whereas the \cite{rugel_x-shooter_2018} and this work values were calculated using a simple slab model. See text for details. $^\dagger$Using $10^4$ erg cm$^{-2}$ s$^{-1}$ as a saturation level (see Figure \ref{fig:civ}), and Equation \ref{eq:macc} from \cite{jk_2000}.}
    \label{tab:maccs}
\end{table*}


The above results suggest that the current $L_\mathrm{CIV}$--$\dot{M}_\mathrm{acc}$ relationship cannot simply be extrapolated down to lower mass objects, and it may overestimate the accretion rate at these low masses. There may be a mass threshold below which this relationship no longer holds, which is important for extending our understanding of accretion into the planetary regime. 

\begin{figure}[hbt!]
    \centering
    \includegraphics[scale=0.35]{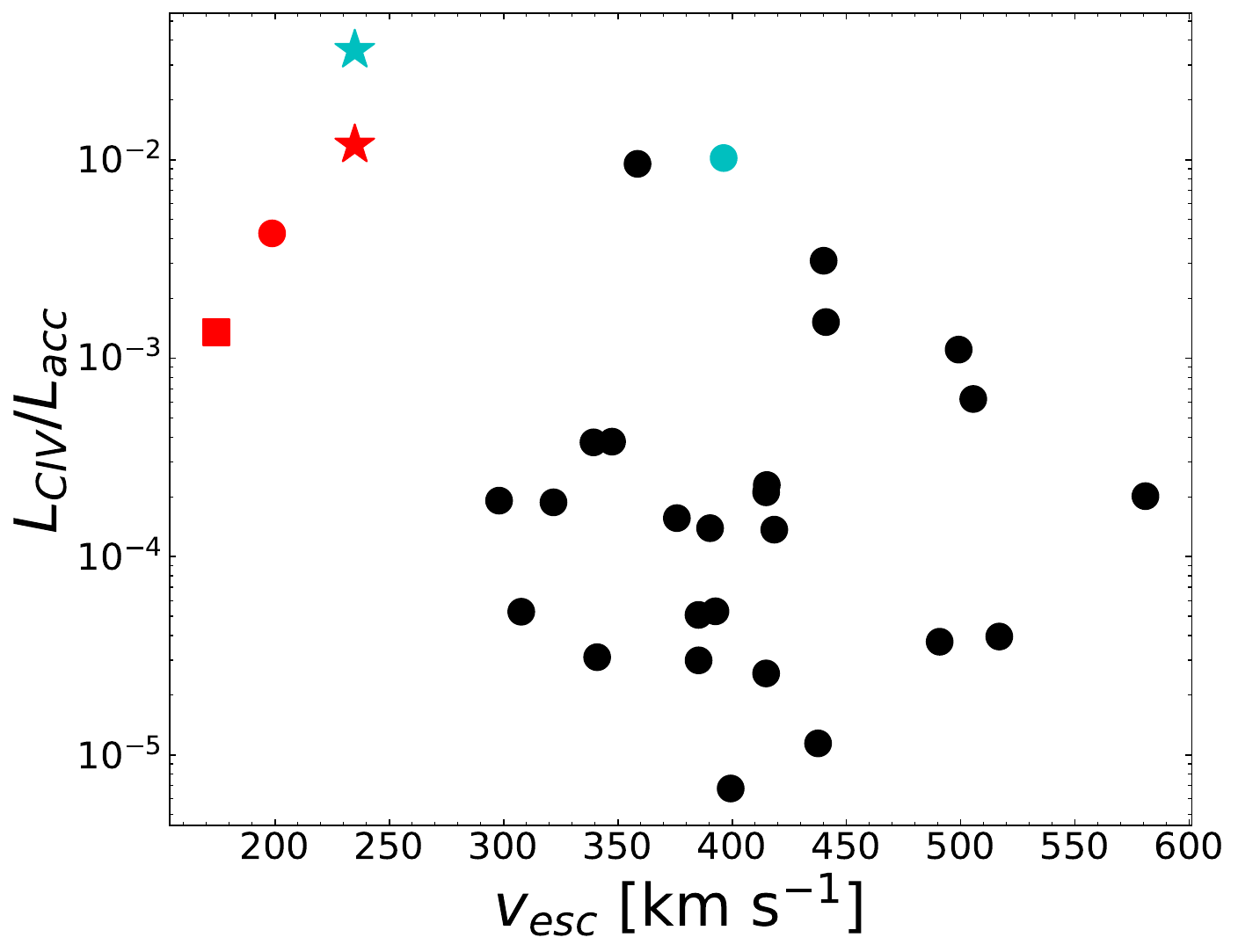}
    \caption{Ratio of C \textsc{\footnotesize{IV}} luminosity ($L_\mathrm{CIV}$) to accretion luminosity ($L_\mathrm{acc}$) vs. escape velocity of the star. The black filled circles are CTTSs. The red points are all brown dwarfs: the red star is J0844, the red filled circle is J1207, and the red square is J0414. The two cyan points are the results of modelling the C \textsc{\footnotesize{IV}} emission of both J0844 (the cyan star) and a ``baseline'' TTS (the cyan filled circle) whose parameters are listed in Table \ref{tab:shockmodparams}.}
    \label{fig:LaccLciv}
\end{figure}

To further explore this relationship, we consider the ratio of the C \textsc{\footnotesize{IV}} luminosity to the overall accretion luminosity. This ratio is expected to probe the accretion shock structure, since C \textsc{\footnotesize{IV}} forms at $\sim$10$^5$ K which is only reached near the shock, while the total accretion luminosity is dominated by the optical continuum which comes from material at $\sim$10$^4$ K. In Figure \ref{fig:LaccLciv}, we compute this ratio for several CTTSs using the values compiled by \cite{yang_far-ultraviolet_2012} (including accounting for extinction), plotted against the escape velocity of each source. The escape velocity is equal to the free-fall velocity if one assumes material falling in from infinity. Assuming material falls in from a certain disc truncation radius would decrease the free-fall velocity reached by the accreting material. However, if we assume $R_\mathrm{in}$ is the same (usually 5$R_*$ is used; \citealt{hartmann_accretion_2016}) then the escape velocity would decrease by the same amount.
The masses and radii of each source are obtained from the CASPER database (\citealt{betti_comprehensive_2023}). The accretion luminosities, also from \cite{yang_far-ultraviolet_2012}, were found by modelling the blue optical continuum excess, and so provide an independent measure of the accretion rate. Although weak, there appears to be a correlation between the $L_\mathrm{CIV}/L_\mathrm{acc}$ ratio and the object's escape velocity: objects with lower escape velocities (i.e., the brown dwarfs, including J0844) show higher $L_\mathrm{CIV}/L_\mathrm{acc}$ ratios compared to objects with higher escape velocities (i.e., the CTTSs). Additionally, \cite{zhou_accretion_2014, zhou2021} find that planetary-mass objects seem to produce hydrogen emission more efficiently (relative to their accretion luminosities) compared to CTTSs. Both of these differences suggest a difference in the shock structure between brown dwarfs and their higher mass CTTS counterparts.

In order to further explore this difference, we compute simple accretion shock models similar to the one described in \cite{calvetgullbring1998} with slight modifications as outlined in \cite{jkardila2003} in order to calculate C \textsc{\footnotesize{IV}} emission. We perform the same calculation for a ``baseline'' TTS. The model assumes an accretion column with one-dimensional, plane-parallel geometry. The shock velocity is assumed to be the free-fall velocity of material falling in from an inner truncation radius of $R_\mathrm{in} = 5$ $R_*$, using 
\begin{equation}
    v_\mathrm{s} = \Biggl(\frac{2 G M_*}{R_*}\Biggr)^{1/2} \Biggl(1 - \frac{R_*}{R_\mathrm{in}} \Biggr)^{1/2}
\end{equation}
The surface area of the accretion footprint is calculated using
\begin{equation}
    A = 4 \pi R_*^2 f
    \label{eq:accfoot}
\end{equation}
where $f$ is the accretion filling factor, i.e., how much of the star is covered by the accretion footprint. The area of the accretion footprint, together with the mass accretion rate and the shock velocity determines the density of the accreting material and the energy flux into the shock, governed by
\begin{equation}
    F = \frac{1}{2} \rho v_\mathrm{s}^3
\end{equation}
We adopt a standard filling factor of 0.02 for both J0844 and the baseline CTTS. A summary of the input parameters for the shock calculations is shown in Table \ref{tab:shockmodparams}.

\begin{table}[hbt!]
    \caption{Parameters used as inputs for the shock model.}
    \begin{center}
    \begin{tabular}{l c c}
    \hline
    Model Parameter & J0844 & Baseline CTTS  \\
    \hline \hline
    $M_*$ ($M_\odot$) & 0.052 & 0.7 \\
    $R_*$ ($R_\odot$) & 0.43 & 1.7 \\
    $\dot{M}_\mathrm{acc}$ ($M_\odot$ yr$^{-1}$) & $4.2 \times 10^{-11}$ & $1 \times 10^{-8}$ \\
    Filling factor ($f$) & $^\dagger$0.02 & 0.02 \\
    \hline
    \end{tabular}
    \end{center}
    \tablecomments{$^\dagger$Standardised to be the same as the baseline CTTS. }
    
    \label{tab:shockmodparams}
\end{table}

The shock model comprises two steps, the first of which takes two input parameters: the energy flux $F$, and the shock velocity $v_\mathrm{s}$, and calculates the gas temperature, mass density and pressure as a function of the distance from the shock front. These parameters are then used as inputs to compute the UV spectrum using the \textsc{\footnotesize{CHIANTI}} database and associated IDL codes (\citealt{dere_chianti_1997}) to calculate the line intensities. The model is described in more detail in \cite{ardila_consistent_2009}. The C \textsc{\footnotesize{IV}} lines produced are then integrated, and multiplied by the surface area of the accretion footprint given by Equation \ref{eq:accfoot} in order to find the C \textsc{\footnotesize{IV}} luminosity ($L_\mathrm{CIV}$). The accretion luminosity ($L_\mathrm{acc}$) is found using Equation \ref{eq:lacc} and the mass accretion rates shown in Table \ref{tab:shockmodparams}.

The model values (cyan points in Figure \ref{fig:LaccLciv}) generally agree well with the observations in terms of their \textit{relative} strengths: the model predicts that J0844 (and therefore possibly brown dwarfs in general) should have a higher $L_\mathrm{CIV}$/$L_\mathrm{acc}$ relationship compared to a ``baseline'' CTTS, which is indeed what we see in the observations. This is further confirmed by J1207's location on the plot (filled red circle). The other brown dwarf, J0414 (filled red square) is less convincing, but still shows a slightly elevated ratio compared to the majority of the CTTSs. However, in both model cases (``baseline'' CTTS and J0844), the model predicts higher $L_\mathrm{CIV}$/$L_\mathrm{acc}$ than what we find. This is likely due to the simplicity of the shock model we are using, which considers only post-shock material assumed to be optically thin. In reality, accretion-related emission (such as C \textsc{\footnotesize{IV}}, which we consider here) must also interact with the pre-shock region which is likely heated by the post-shock gas and the opacity in both regions should be accounted for. This highlights the need for more detailed modelling of these quantities. In Figure \ref{fig:ratio} and \ref{fig:ratio_age}, we only show the ``baseline'' CTTS model (cyan filled circle) because the C/Si ratio does not change when using the shock model for J0844 because both \textsc{\footnotesize{IV}} and Si \textsc{\footnotesize{IV}} trace gas at the same temperature ($10^5$ K).


\section{Summary} \label{sec:summary}

We have presented FUV and optical \textit{HST} observations of the young accreting brown dwarf, J0844. Our principal results are:

\begin{itemize}
    \item We find that J0844's C \textsc{\footnotesize{IV}} emission profiles are more similar to WTTSs than CTTSs in terms of their widths, despite J0844 showing evidence for ongoing accretion activity. 
    \item We estimate a C/Si ratio that is higher than that measured in
    most CTTSs and WTTSs, indicating that at least some of the Si is locked up in grains, though not as much as for some CTTSs, like TW Hya, which have a larger ratio than J0844.  We find that the C/Si ratio correlates
    somewhat with stellar mass and more strongly with stellar age in
    agreement with recent theoretical studies of \citet{mah2023} and
    \citet{huhn2023}.
    \item We calculate an accretion rate of $4.2 \times 10^{-11}$ $M_\odot$ yr$^{-1}$ by fitting the optical continuum with a simple hydrogen slab model, which is about two thirds of what \cite{rugel_x-shooter_2018} found. If we compare J0844 to a similar brown dwarf, J1207, the accretion rate we calculate for J0844 is significantly lower than what was found for J1207 by \cite{france_metal_2010}, despite J0844 showing higher levels of C \textsc{\footnotesize{IV}} flux. \cite{france_metal_2010} extrapolated the $L_\mathrm{CIV}$--$\dot{M}_\mathrm{acc}$ for higher mass stars to determine $\dot{M}_\mathrm{acc}$ for J1207, and we suggest that this relationship may not be able to be extended down into the lowest mass regime, and may overestimate the accretion rate.  
    \item We find that the $L_{\mathrm{CIV}}/L_{\mathrm{acc}}$ ratio is larger for J0844, J1207 and J0414 compared to most other CTTSs, which we confirm through a simple shock model. This may explain part of the discrepancy between estimating the accretion rate using the optical continuum versus excess C \textsc{\footnotesize{IV}} emission for brown dwarfs. This suggests that using excess C \textsc{\footnotesize{IV}} may systematically overestimate the accretion rate in very low mass objects like brown dwarfs and giant planets. It also suggests that the accretion shock structure might be different in brown dwarfs compared to CTTSs.
    \item We fit the observed H$_2$ emission lines in two ways: using a double Gaussian, then using a stellar plus disc profile. Much of the H$_2$ emission appears to arise from inside the co-rotation radius ($R_\mathrm{co} \approx 4.9$ $R_*$); we recover inner emission radii of $1.9$ $R_*$ using the double Gaussian fit, and $3.3$ $R_*$ using a more realistic disc model. This suggests the inner disc is truncated inside the co-rotation radius, with our best estimate that the disc is truncated at $0.7$ $R_\mathrm{co}$ which is consistent with observed inner truncation radii for many CTTSs \citep{salyk_co_2011}.
\end{itemize}
The above findings suggest that the accretion process in brown dwarfs might be somewhat different compared to their higher mass counterparts. Analogously, this may also be true at even lower masses such as giant planets. Therefore, characterising the accretion flows of brown dwarfs with UV studies is important in extending our knowledge of accretion down to the lowest masses, and thus fully bridging the gap between stars and planets. A better understanding of accretion processes in brown dwarfs will provide important insights into interpreting observations of forming giant planets.

\begin{center}
    ACKNOWLEDGMENTS
\end{center}
\begin{acknowledgments}
The implementation of the ULLYSES project would not have been successful without the work of Will Fischer, who passed away in 2024. We recognise his immense contribution to STScI, this project, and the field in general.

We thank ODYSSEUS collaboration members Caeley Pittman, Kevin France, and Silvia Alencar for insightful feedback that improved the manuscript. We also thank the anonymous referee for helpful suggestions.

This work was supported by grant HST AR-16129.014 to Rice University and benefited from discussions with the ODYSSEUS team (https://sites.bu.edu/odysseus/; \citealt{espaillat_odysseus_2022}).  We also wish to acknowledge additional funding from NASA through award 80-NSSC18K-0828 made to Rice University.

Observations were obtained with NASA/ESA Hubble Space Telescope. Data were retrieved from the Mikulski Archive for Space Telescopes (MAST) at the Space Telescope Science Institute (STScI), operated by the Association of Universities for Research in Astronomy, Inc. under NASA contract NAS 5-26555. The ULLYSES data are from \cite{https://doi.org/10.17909/t9-jzeh-xy14}.

This work also made use of observations obtained by the European Southern Observatory with VLT/X-Shooter as part of the PENELLOPE programme \citep{manara_2023_10024001}. We thank Laura Venuti for the reduction of these data.

This paper includes data collected by the TESS mission. Funding for the TESS mission is provided by the NASA's Science Mission Directorate.

Based on observations collected at the European Southern Observatory under ESO programme 106.20Z8 and on data obtained from the ESO Science Archive Facility \citep{https://doi.org/10.18727/archive/88} under Program IDs 090.C-0050(A), 093.C-0658(A), 094.C-0805(A), and 108.22M8.001.


\end{acknowledgments}

%

\vspace{5mm}
\facilities{HST (COS, STIS), VLT (X-Shooter), TESS}


\software{\textsc{numpy} (\citealt{harris2020array}), \textsc{pandas} (\citealt{reback2020pandas, mckinney-proc-scipy-2010}), \textsc{matplotlib} (\citealt{matplotlib}), \textsc{astropy} (\citealt{2013A&A...558A..33A,2018AJ....156..123A}), \textsc{scipy} (\citealt{2020SciPy-NMeth}), \textsc{lightkurve} (\citealt{2018ascl.soft12013L}), \textsc{emcee} (\citealt{emcee})}





\bibliography{sample631}{}
\bibliographystyle{aasjournal}



\end{document}